\documentclass[a4paper,fleqn,usenatbib]{mnras}
\usepackage{newtxtext,newtxmath}
\usepackage[T1]{fontenc}
\usepackage{ae,aecompl}
\usepackage{graphicx}	% Including figure files
\usepackage{amsmath}	% Advanced maths commands
\usepackage{amssymb}	% Extra maths symbols
\usepackage{color}	
\newcommand{\Msun}{\hbox{$\rm\thinspace M_{\odot}$}}

\newcommand{\gtsim}{\mbox{{\raisebox{-0.4ex}{$\stackrel{>}{{\scriptstyle\sim}}$}}}}

\title[The VANDELS spectroscopic survey]{The VANDELS ESO public spectroscopic survey}
\author [R. J. McLure et al.]{R. J. McLure$^{1}$, L. Pentericci$^{2}$, A. Cimatti$^{3,4}$, J. S. Dunlop$^{1}$, D. Elbaz$^{5}$, A. Fontana$^{2}$, 
\newauthor K. Nandra$^{6}$, R. Amorin$^{7,8}$, M. Bolzonella$^{9}$, A. Bongiorno$^{2}$, A. C. Carnall$^{1}$, 
\newauthor M. Castellano$^{2}$, M. Cirasuolo$^{10}$, O. Cucciati$^{9}$, F. Cullen$^{1}$, S. De Barros$^{11}$, 
\newauthor S. L. Finkelstein$^{12}$, F. Fontanot$^{13}$, P. Franzetti$^{14}$, M. Fumana$^{14}$,  A. Gargiulo$^{14}$,
\newauthor B. Garilli$^{14}$, L. Guaita$^{2,15}$, W. G. Hartley$^{16}$, A. Iovino$^{17}$, M. J. Jarvis$^{18}$, S. Juneau$^{5}$,
\newauthor W. Karman$^{19}$, D. Maccagni$^{14}$, F. Marchi$^{2}$, E. M\'armol-Queralt\'o$^{1}$, E. Pompei$^{20}$,
\newauthor L. Pozzetti$^{9}$, M. Scodeggio$^{14}$, V. Sommariva$^{3}$, M. Talia$^{3,9}$, O. Almaini$^{21}$, I. Balestra$^{22}$, 
\newauthor S. Bardelli$^{9}$, E. F. Bell$^{23}$, N. Bourne$^{1}$, R. A. A. Bowler$^{18}$, M. Brusa$^{3}$, F. Buitrago$^{24,25}$,
\newauthor K. I. Caputi$^{19}$, P. Cassata$^{26}$, S. Charlot$^{27}$, A. Citro$^{3}$, G. Cresci$^{4}$, S. Cristiani$^{13}$, 
\newauthor E. Curtis-Lake$^{27}$, M. Dickinson$^{28}$, G. G. Fazio$^{29}$, H. C. Ferguson$^{30}$, F. Fiore$^{2}$, 
\newauthor M. Franco$^{5}$,  J. P. U. Fynbo$^{31}$, A. Galametz$^{6}$, A. Georgakakis$^{32}$, M. Giavalisco$^{33}$,
\newauthor A. Grazian$^{2}$, N. P. Hathi$^{30}$, I. Jung$^{12}$, S. Kim$^{34}$, A. M. Koekemoer$^{30}$, Y. Khusanova$^{35}$,
\newauthor O. Le F\`evre$^{35}$, J. M. Lotz$^{30}$, F. Mannucci$^{4}$, D. T. Maltby$^{21}$, K. Matsuoka$^{4}$,
\newauthor D. J. McLeod$^{1}$, H. Mendez-Hernandez$^{26}$, J. Mendez-Abreu$^{36, 37}$, M. Mignoli$^{3}$,
\newauthor M. Moresco$^{3,9}$, A. Mortlock$^{1}$, M. Nonino$^{13}$, M. Pannella$^{38}$, C. Papovich$^{39}$, P. Popesso$^{40}$,
\newauthor D. P. Rosario$^{41}$, M. Salvato$^{6, 40}$, P. Santini$^{2}$, D. Schaerer$^{11}$, C. Schreiber$^{42}$, D. P. Stark$^{43}$,
\newauthor L. A. M. Tasca$^{35}$, R. Thomas$^{20}$, T. Treu$^{44}$, E. Vanzella$^{9}$, V. Wild$^{45}$, C. C. Williams$^{43}$,
\newauthor G. Zamorani$^{9}$, E. Zucca$^{9}$\\
\\
Affiliations are listed at the end of the paper}

\date{Accepted XXX. Received YYY; in original form ZZZ}
\pubyear{2018}
\begin{document}
\label{firstpage}
\pagerange{\pageref{firstpage}--\pageref{lastpage}}

\maketitle

\begin{abstract}
VANDELS is a uniquely-deep spectroscopic survey of high-redshift
galaxies with the VIMOS spectrograph on ESO's Very Large Telescope (VLT).
The survey has obtained ultra-deep optical ($0.48 < \lambda < 1.0$ {\rm $\mu$}m) spectroscopy of $\simeq$2100 galaxies within the redshift
interval $1.0\leq z \leq7.0$, over a total area of $\simeq 0.2$ deg$^2$ centred
on the CANDELS UDS and CDFS fields. Based on accurate photometric
redshift pre-selection, 85\% of the galaxies targeted by VANDELS were
selected to be at $z\geq3$.
Exploiting the red sensitivity of the refurbished VIMOS spectrograph,
the fundamental aim of the survey is to provide the 
high signal-to-noise ratio spectra necessary to measure key physical properties such as stellar 
population ages, masses, metallicities and outflow velocities from detailed absorption-line studies. 
Using integration times calculated to produce an approximately constant
signal-to-noise ratio ($20~<~t_{\rm int}<~80$~hours), the VANDELS 
survey targeted: {\it a)}  bright star-forming galaxies at $2.4 \leq z
\leq 5.5$, {\it b)}  massive quiescent galaxies at
$1.0 \leq z \leq 2.5$, {\it c)} fainter star-forming galaxies at $3.0
\leq z \leq 7.0$ and {\it d)} X-ray/{\it Spitzer-}selected active
galactic nuclei and {\it Herschel}-detected galaxies. By targeting two extragalactic survey fields
with superb multi-wavelength imaging data, VANDELS will produce a
unique legacy data set for exploring the physics underpinning
high-redshift galaxy evolution. In this paper we provide an overview of the VANDELS survey designed to
support the science exploitation of the first ESO public data release,
focusing on the scientific motivation, survey design and target selection.
\end{abstract}


\section{Introduction}
Understanding the formation and evolution of galaxies remains the key goal
of extra-galactic astronomy. However, delineating the evolution of galaxies,
from the collapse of the first gas clouds at early times to the
assembly of the complex structure we observe in the local Universe,
continues to present an immense observational (e.g. \citealt{mad14}) and theoretical challenge (e.g. \citealt{somer15}; \citealt{knebe15}). 

From an observational perspective, the last fifteen years have been a
period of unprecedented progress in our understanding of the basic
demographics of high-redshift galaxies. As a direct consequence of the
availability of deep, multi-wavelength, survey fields, we now have a
good working knowledge of how the galaxy luminosity function
(e.g. \citealt{mclure13}; \citealt{bowler15}; \citealt{fink16}; \citealt{mort17}), stellar mass function 
(e.g. \citealt{muzzin13}; \citealt{tom14}; \citealt{david17}) and global star-formation rate density (SFRD) evolve with redshift
(e.g. \citealt{mag13}; \citealt{novak17}).
Indeed, \cite{mad14} recently demonstrated the consistency (within a factor of $\sim2$) between the
integral of current SFRD determinations and direct estimates of the evolution of stellar-mass density. 

As a consequence, we can now be confident that the low SFRD we
observe locally is approximately the same as it was when the Universe
was less than 1 Gyr old (i.e. $z\simeq 7$), and that in the
intervening period the Universe was forming stars up to $\geq10$ times more rapidly.
However, despite this, it is still perfectly plausible to argue that the peak in cosmic star-formation 
occurred anywhere in the redshift interval $1.5<z<3.5$, an
uncertainty of 2.5 Gyr. Moreover, the results of the latest generation of semi-analytic and hydro-dynamical galaxy 
simulations (e.g. \citealt{genel14}; \citealt{henri15}; \citealt{somer15}) demonstrate that, from a
theoretical perspective, even reproducing the evolution of the cosmic
SFRD can still be problematic.

Over the last decade it has become established that the majority of
cosmic star formation is produced by galaxies lying on the so-called `main
sequence' of star formation (\citealt{noeske07}; \citealt{elbaz07}; \citealt{daddi07}), 
a roughly linear relationship between star-formation rate (SFR) and stellar mass, the normalisation of which increases with look-back time.
Furthermore, the evolving normalisation of the main sequence over the last
10 Gyr is now relatively well determined, with the average SFR at a given stellar mass increasing by a factor of $\simeq
30$ between the local Universe and redshift $z\simeq 2$ (e.g. \citealt{whitaker14}; \citealt{speagle14}; \citealt{johnston15}).
However, at higher redshifts the evolution of the main sequence is still uncertain, despite a clear theoretical
prediction that it should mirror the increase in halo gas accretion
rates (i.e. $\propto (1+z)^{2.5}$; \citealt{dekel09}). Depending on
their assumptions regarding star-formation histories, metallicity,
dust and nebular emission, different studies find that 
the increase in average SFR between $z=2$ and $z=6$ at a given stellar
mass is anything from a factor of $\simeq 2$ (e.g. \citealt{gonz14}; \citealt{esther16})
to a factor of $\simeq 25$ (e.g. \citealt{barros14}); see \cite{stark16} for a recent review.

Although the decline of the global SFRD at $z\leq 2$ is now well characterised observationally, the relative
importance of the different physical drivers responsible for the quenching of star formation remains unclear.
With varying degrees of hard evidence and speculation, feedback from
active galactic nuclei (AGN), stellar winds, merging and environmental/mass
driven quenching have all been widely discussed in the recent
literature (e.g. \citealt{fabian12}; \citealt{conselice14}; \citealt{peng15}).
At some level, quenching must be connected to the interplay between gas outflow, the inflow of
`pristine' gas  and morphological transformation. However, to date, the precise roles played by the different underlying
physical mechanisms still remain uncertain, as does the potential redshift
evolution of the quenching process.
Indeed, recent evidence based on deep optical and near-IR
spectroscopy strongly suggests that the physical properties of 
star-forming galaxies at $z=2-3$ are significantly different from their low-redshift counterparts in
terms of metallicity, $\alpha-$enhancement and ionization parameter
(e.g. \citealt{ferg14}; \citealt{shap15}; \citealt{steidel16};
\citealt{ferg16}; \citealt{strom17}). Moreover, recent results at
sub-mm and mm-wavelengths with {\it Herschel} and ALMA indicate that 
the dust properties of star-forming galaxies at high redshift may also
be significantly different (e.g. \citealt{capak15}; \citealt{bow16}; \citealt{Reddy17}), although
the current picture is far from clear (e.g. \citealt{d17};
\citealt{nathan17}; \citealt{mclure17}; \citealt{kop18}; \citealt{bowler18}). 

In summary, it now appears that progress in our
understanding of galaxy evolution at high redshift is often less
limited by poor statistics than by the systematic uncertainties in our
measurements of the crucial physical parameters, caused by the insidious and interrelated degeneracies between age, dust attenuation and metallicity.
It is also clear that substantive progress in addressing these
uncertainties will rely on combining the best available
multi-wavelength imaging with deep spectroscopy (e.g. \citealt{kurk13}).
Within this context, a series of spectroscopic campaigns with
VLT+VIMOS, such as the VIMOS Very Deep Survey (VVDS;
\citealt{lefevre05}), the COSMOS spectroscopic survey (zCOSMOS;
\citealt{lilly07}) and the VIMOS Ultra Deep Survey (VUDS;
\citealt{lefevre15}), have played a key role in improving our
understanding of galaxy evolution, primarily through providing 
large numbers of spectroscopic redshifts over wide fields. 
The VANDELS survey is designed to complement and extend the 
work of these previous campaigns by focusing on ultra-long exposures
of a relatively small number of galaxies, pre-selected to lie at high redshift using the best available photometric redshift information.

The VANDELS survey is a major new ESO Public Spectroscopic Survey using the VIMOS spectrograph on the VLT to obtain ultra-deep,
medium resolution, red-optical spectra of $\simeq 2100$ high-redshift galaxies. 
The survey was allocated 914 hours of VIMOS integration time and,
between August 2015 and February 2018, each target galaxy received
20--80 hours of on-source integration, obtained via repeated observations of the UDS and CDFS multi-wavelength survey fields.
The fundamental science goal of VANDELS is to move beyond
redshift acquisition and obtain a spectroscopic data set deep enough 
to study the astrophysics of high-redshift galaxy evolution.
The VANDELS spectroscopic targets were all pre-selected using
high-quality photometric redshifts, with the vast majority ($\simeq
97\%$) drawn from three main categories. Firstly, VANDELS targeted bright ($i_{\rm AB}\leq 25$) star-forming galaxies in the redshift
range $2.4 \leq z \leq 5.5$ (median $z=2.8$). For these galaxies, the
signal-to-noise ratio (SNR) and wavelength coverage of the VANDELS spectra are designed to allow stellar metallicity and gas 
outflow information to be extracted for individual objects.
Secondly, to study the descendants of high-redshift star-forming
galaxies, VANDELS targeted a complementary sample of massive
($H_{\rm AB}\leq 22.5$) passive galaxies at $1.0 \leq z \leq 2.5$ (median $z=1.2$). 
Again, in combination with deep multi-wavelength photometry and 3D-HST
grism spectroscopy \citep{brammer12}, the high SNR spectra provided by
VANDELS are designed to provide age/metallicity information and
star-formation history constraints for individual objects.
Thirdly, VANDELS extended to fainter magnitudes and
higher redshifts by targeting a large statistical sample of faint
star-forming galaxies ($25\leq H_{\rm AB}\leq 27$, $i_{\rm AB}\leq 27.5$) in
the redshift range $3 \leq z \leq 7$ (median $z=3.5$). 
Throughout the rest of the paper we will refer to the galaxies in this
sample as Lyman-break galaxies (LBGs), although they were not selected
via traditional colour-colour criteria (see Section 4). The final $\simeq 3\%$ of VANDELS spectroscopic slits were allocated
to AGN candidates or {\it Herschel-}detected galaxies with $i_{\rm
  AB}\leq 27.5$ and $z \geq 2.4$ (median $z=2.7$).

In this paper we provide an overview of the VANDELS survey to support
the science exploitation of the first data release (DR1) via the ESO
Science Archive Facility ({\tt archive.eso.org}). The structure of the paper is as follows. 
In Section 2 we provide a brief review of the science cases that provided the principal motivation for VANDELS, along with the multiple
legacy science cases which could be facilitated by the data. In Section 3 we describe the reasoning behind the choice of survey
fields. In Section 4 we describe the target selection process, including the
generation of photometric catalogues and the determination of robust
photometric redshifts. In Section 5 we describe the basic observing
strategy before providing brief details of the data reduction and
spectroscopic redshift measurement procedures in Section 6. In Section 7 we describe the contents of the first data release, before reviewing the
success of the VANDELS target selection process using the on-sky DR1 data in Section 8.
A full description of DR1, including a detailed discussion of the observing strategy, data reduction and spectroscopic redshift
measurements is provided in a companion data release paper \citep{dr1paper}. In Section 9 we provide a summary and an overview of the content and
timeline for subsequent data releases. Throughout the
paper we refer to total magnitudes quoted in the AB system 
\citep{oke83}. We assume the following cosmology: $\Omega_{M}=0.3$,
$\Omega_{\Lambda}=0.7$ and $H_{0}=70$ km~s$^{-1}$~Mpc$^{-1}$, and adopt
a \cite{chab03} initial mass function (IMF) for calculating stellar masses and star-formation rates.

\section{Science Motivation}
The primary motivation behind the VANDELS survey was to provide
spectra of high-redshift galaxies with sufficiently high SNR to
allow absorption line studies both on individual objects and via stacking.
Armed with spectra of sufficient quality it should be possible, in combination with excellent multi-wavelength photometry, to provide
significantly improved constraints on key physical parameters such as stellar mass, star-formation rate, metallicity and dust attenuation.
As a result, it is clear that the data set provided by VANDELS will
have a potentially significant impact on many different areas of
high-redshift galaxy evolution science. In this section we provide a
concise overview of the key science goals that motivated the original
VANDELS survey proposal, before briefly reviewing the legacy science case.

\subsection{Stellar metallicity and dust attenuation}
Tracing the evolution of metallicity is a powerful method of
constraining high-redshift galaxy evolution, due to its direct link to past star formation and sensitivity to interaction
(i.e. gas inflow/outflow) with the inter-galactic medium (e.g. \citealt{mannucci10}).
Moreover, accurate knowledge of metallicity is essential for deriving accurate
star-formation rates and breaking the degeneracy between age and dust attenuation (e.g. \citealt{sandy14}).
Consequently, it is clear that extracting constraints on the metallicity and dust attenuation of 
high-redshift galaxies from VANDELS spectra is important to investigations of the build-up of the {\it  stellar}
mass-metallicity relation, accurately quantifying the peak in cosmic
star-formation history (e.g. \citealt{cast14}; \citealt{d17}), and resolving the
current uncertainties regarding the evolution of sSFR at $z\geq 2$ (e.g. \citealt{stark16}).

Recent studies using stacked spectra of relatively small samples (e.g. \citealt{steidel16}) have shown that is possible to derive
accurate stellar metallicities from the rest-frame UV spectra of
galaxies at $z\geq 2$, given a sufficiently high SNR.
In addition, \cite{steidel16} also demonstrated that rest-frame UV
spectra can potentially be used to quantify the impact of binary stars in
stellar population synthesis models (e.g. \citealt{stan16}; \citealt{jj16}) by
fitting to the He{\sc ii} emission line at $1640$\AA. 

The high SNR and accurate flux calibration of the VANDELS
spectra facilitates the measurement of stellar metallicities using photospheric UV absorption lines ($1370-1900$\AA),
whose equivalent width is sensitive to metallicity and independent of
other stellar parameters (e.g. \citealt{som12}; \citealt{rix04}). 
Moreover, within the context of dust attenuation, the VANDELS data set
also has the potential to differentiate between competing dust
reddening laws (e.g. \citealt{ferg18}; \citealt{mclure17}), and to
constrain the strength of the 2175\AA\, bump.

The final VANDELS data set will provide individual and stacked
measurements of stellar metallicity based on $\gtsim1000$ spectroscopically-confirmed
star-forming galaxies in the redshift range $2.4 \leq z \leq 5.0$. These measurements can be compared with the
gas-phase metallicities currently being derived for $z\simeq 2.5$
galaxies by the MOSDEF \citep{shap15} and KBSS-MOSFIRE \citep{strom17}
surveys and forthcoming observations with the {\it James Webb Space Telescope} ({\it JWST}).

\subsection{Outflows}
Along with stellar-metallicity measurements, a key science goal for
VANDELS is to investigate the role of stellar and AGN feedback in
quenching star formation at high redshift via studies of outflowing
interstellar gas. Over recent years it has become established that high-velocity
outflows are likely to be ubiquitous for star forming galaxies at $z>1$
(e.g. \citealt{weiner09}), with mass outflow rates comparable to the
rates of star formation (e.g. \citealt{brad13}), and that very compact
starbursts can produce outflows with velocities $>1000$~km~s$^{-1}$,
yielding winds that were previously only thought possible from AGN
activity \citep{diamond12}. It seems likely that such outflows are playing a major role in the termination of star formation
at high redshift and the build-up of the mass-metallicity relation.

The individual and stacked spectra of star-forming
galaxies delivered by VANDELS will provide accurate measurements of outflowing ISM velocities from high and
low-ionization UV interstellar absorption features (e.g. \citealt{shap03}), allowing the outflow rate to be
investigated as a function of stellar mass, SFR and galaxy
morphology. This offers the prospect of improving our understanding of the impact of galactic outflows on star-formation at $z\geq
2$, directly testing models of the evolving gas reservoir
(e.g. \citealt{dayal13}) and addressing the origins of the Fundamental Mass-Metallicity 
Relation \citep{mannucci10}. Finally, comparing the outflow
velocities of star-forming galaxies with and without hidden AGN
(e.g. \citealt{talia16}) will allow the role of AGN feedback in
quenching star formation and the build-up of the red sequence to be investigated (e.g. \citealt{cimatti13}).

\subsection{Massive galaxy assembly and quenching}
A key sub-component of the VANDELS survey was obtaining deep
spectroscopy of $>250$ massive, passive galaxies at $1.0 \leq z \leq 2.5$. 
This population holds the key to understanding the quenching mechanisms responsible for producing the strong colour bi-modality
observed at $z<1$, together with the significant evolution in the
number density, morphology and size of passive galaxies observed between $z=2$ and the
present day (e.g. \citealt{bruce12}; \citealt{pearce13};
\citealt{tom14}; \citealt{arjen14}). The physical parameters which will be delivered by the VANDELS spectra
offer the prospect of connecting these quenched galaxies with
their star-forming progenitors at $z\geq 3$ in a self-consistent way.

For the majority of the passive sub-sample, the VANDELS spectra provide a combination of crucial 
rest-frame UV absorption-line information (e.g. MgUV,
2640\AA/2900\AA\,breaks) and Balmer-break measurements.
Combined with the unrivalled photometric data available in the UDS and
CDFS fields, it will be possible to break age/dust/metallicity degeneracies and deliver
accurate stellar mass, dynamical mass, star-formation rate,
metallicity and age measurements via full spectrophotometric SED
fitting (e.g. \citealt{pearce13}; \citealt{beagle}; \citealt{carnall18}).

\begin{figure*}
\begin{center}
\includegraphics[angle=0,width=\textwidth]{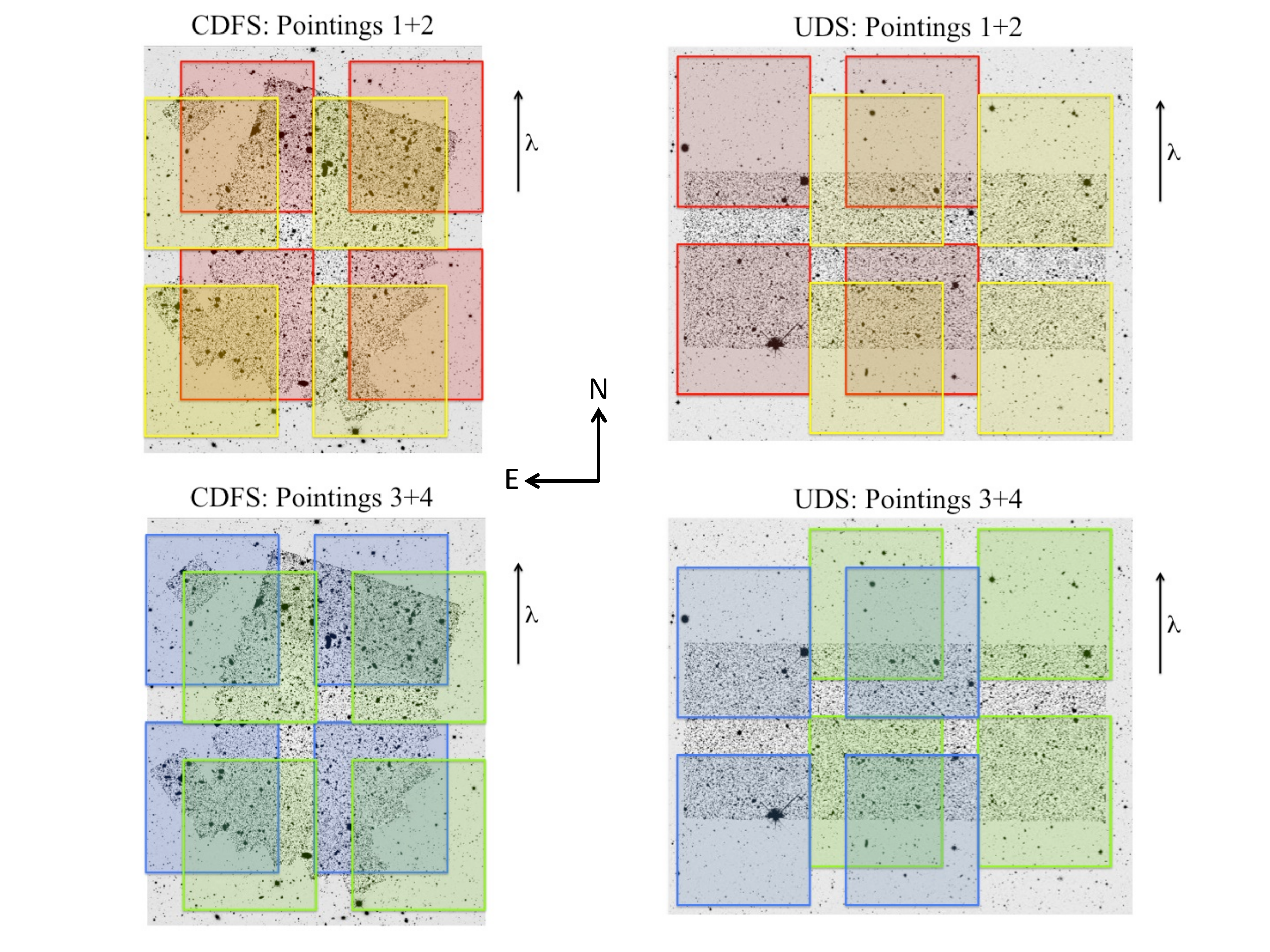}
\caption{Layout of the eight VANDELS pointings,  four in UDS and four
  in CDFS. In each figure the VIMOS quadrants of a given pointing are shown as a different colour, overlaid on a greyscale
  image showing the {\it HST} $H-$band imaging provided by the
  CANDELS survey (\citealt{grogin11}; \citealt{koke11}) in the central regions and ground-based $H-$band
  imaging from the UKIDSS UDS (Almaini et al., in preparation) and
  VISTA VIDEO (\protect\citealt{jarvis13}) surveys covering the wider
  fields. The total area covered by the eight VIMOS pointings is $\simeq 0.2$ square degrees.
The spectroscopic slits are all placed E-W on the sky, as recommended
to minimise slit losses during long VIMOS integrations on fields at these declinations \citep{sanchez}.}
\end{center}
\end{figure*}

\subsection{Legacy Science}
Although the science cases outlined above provided the primary
motivation, as an ESO public spectroscopy survey, the greatest strength
of VANDELS is arguably its long-term legacy value to the astronomical community.
In general terms, by providing high SNR continuum
spectroscopy of galaxies which traditionally only have Ly$\alpha$
redshifts at best, VANDELS is guaranteed to open up new parameter space for
investigating the physical properties of high-redshift galaxies.

More specifically, the VANDELS spectra provide the opportunity to
accurately determine the fraction of Ly$\alpha$ emitters
amongst the general Lyman-break galaxy population in the redshift
range $3.0<z<6.0$, thereby providing an
improved baseline measurement for studies within the reionization epoch
(e.g. \citealt{lake12}; \citealt{pent14}; \citealt{debarros17}). In
addition, VANDELS will also provide large samples 
of spectroscopically-confirmed galaxies at $z\simeq 3$ with which to
identify and study Lyman continuum emitters
(e.g. \citealt{vanzella16}; \citealt{debarros16}; \citealt{shapley16};
\citealt{marchi17}). Moreover, combining the VANDELS spectra with near-IR spectroscopy
offers the prospect of directly comparing stellar and gas-phase
metallicities out to $z\simeq 3.5$, and constraining the possible
star-formation timescales via quantifying the level of $\alpha-$enhancement (e.g. \citealt{steidel16}) as a function of stellar mass and star-formation rate.
We also note that additional science will be facilitated by the samples of rarer 
{\it Herschel-}detected galaxies and AGN targeted by VANDELS. For
these systems, the deep VANDELS spectroscopy will make it possible to assess their physical conditions
(e.g. metallicities, ionizing fluxes and outflow signatures) and compare them with those of less active systems at the same redshifts.

In terms of future follow-up observations, there is an excellent
synergy between VANDELS and the expected launch date of the {\it JWST}
in 2020. The opportunity to combine ultra-deep optical spectroscopy with the unparalleled near-IR spectroscopic
capabilities of NIRSpec will make VANDELS sources an obvious
choice for follow-up spectroscopy with {\it JWST}. For high multi-plex
follow-up observations, there is also an excellent overlap between the footprint of the
VANDELS survey within the UDS and CDFS fields and the the field of
view of ESO's forthcoming Multi Object Optical and Near-infrared Spectrograph (MOONS) for the VLT \citep{moons14}.

Finally, it is also worth noting that the declinations of the UDS and
CDFS fields make them ideal for sub-mm and mm follow-up observations with ALMA. One of the key
scientific questions that VANDELS will help to address is the
evolution of star formation and metallicity in
galaxies at $z\geq 2$. However, in order to derive a complete picture
it will be necessary to obtain dust mass and star-formation rate measurements at long wavelengths, which can now be provided by short, 
targeted, continuum observations with ALMA. 

\section{Field Choice}
The VANDELS survey targets two fields, the UKIDSS Ultra Deep Survey
(UDS: 02:17:38, $-$05:11:55) and the Chandra Deep Field South (CDFS: 03:32:30, $-$27:48:28). Both fields were selected on the basis of their
observability from Paranal and the quality of their existing multi-wavelength ancillary data. 
We note that the COSMOS field, which was also actively
considered for inclusion in VANDELS, was targeted with VIMOS by the ESO public spectroscopy survey LEGA-C \citep{legac}.

Both the UDS and CDFS offer deep optical-nearIR {\it HST}
imaging provided by the CANDELS survey (\citealt{grogin11}; \citealt{koke11}) with the CDFS
also offering deep {\it HST}/ACS optical imaging from the original
GOODS survey \citep{goods} and ultra-deep X-ray imaging \citep{luo17}. Moreover, both fields feature the 
deepest available {\it Spitzer} IRAC imaging on these angular scales
from the S-CANDELS survey \citep{scandels} and deep WFC3/IR grism spectroscopy from the public 3D-HST
programme \citep{brammer12}. When combined with the deepest
available $Y+K$ imaging from the HUGS survey \citep{font14}, it is
clear that the UDS and CDFS are excellent legacy fields for studying the high-redshift Universe. 

Given that a single pointing of the VIMOS spectrograph covers an area
larger than the {\it HST} imaging in any of the five CANDELS fields, another important consideration when choosing which fields to target
with VANDELS was the quality of the ancillary data over a wider area. The importance of the wider-field ancillary data can be seen from
Fig.~1, which shows the layout of the eight VIMOS pointings targeted by the VANDELS survey in UDS and CDFS.
It can be seen that, although the VIMOS pointings are
arranged to ensure that all of the deep WFC3/IR imaging is
covered, approximately 50\% of the full VANDELS survey footprint lies
outside the central areas of the UDS and CDFS fields that are covered by {\it HST} imaging. 
Crucially, in both the UDS and CDFS, these wider-field regions are covered by
high-quality, publicly-available, optical-nearIR imaging data from a wide variety of different
ground-based telescopes (see Table 1). 
\begin{table}
\caption{Details of the imaging data included in the new photometric catalogues
  generated for the wide-field areas of the CDFS and UDS fields. Column 1 lists the field,
  column 2 lists the filters, column three lists the median 5$\sigma$
  depths measured within a $2^{\prime\prime}-$diameter aperture, 
column 4 lists the telescopes on which the imaging was
  obtained and column 5 lists the paper where the data are presented.
  For the two filters tagged with a $\dagger$ in
  column 2, the $5\sigma$ depth refers to the depth measured after the
  {\it HST} imaging was convolved to match the
  1.0$^{\prime\prime}$ FWHM spatial resolution of the
  ground-based imaging in CDFS. The filters listed as `IA' in
  column 2 are medium-band filters and NB921 is a narrow-band
  filter. The two $z-$band filters listed for the UDS field
  (z$_{1}^{\prime}$ and z$_{2}^{\prime}$) refer to imaging obtained
  with the Suprime-Cam z$^{\prime}-$filter before and after the CCD
  detectors were upgraded. The references listed in column 5
  correspond to: (1) Almaini et al., in preparation, (2)
  \protect\cite{furusawa08}, (3) \protect\cite{furusawa16}, (4)
  \protect\cite{sobral12}, (5) \protect\cite{jarvis13}, (6) \protect\cite{non09}, (7) \protect\cite{card10},
  (8) \protect\cite{rix04}, (9) \protect\cite{tenis12}.}
\begin{center}
\begin{tabular}{ccccc}
\hline
Field & Filter & Depth($5\sigma$) & Telescope & Reference\\
\hline
\hline
UDS & U& 27.0 & CFHT&1\\

       & B& 27.8 & Subaru&2\\

       & V& 27.4 & Subaru&2\\

       & R& 27.2 & Subaru&2\\

       & i$^{\prime}$& 27.0 & Subaru&2\\

       & z$_{1}^{\prime}$& 26.0 & Subaru&2\\

       & z$_{2}^{\prime}$& 26.4 & Subaru&3\\

       & NB921& 25.8 & Subaru&4\\

       & Y         &25.1& VISTA&5\\

       & J          &25.5& UKIRT&1\\

       & H         &24.9& UKIRT&1\\

       & K         &25.1& UKIRT&1\\
\hline
CDFS & U         & 27.8 & VLT &6\\

         & B          & 27.1 & ESO 2.2m&7\\

         & IA484  & 26.4 & Subaru&7\\

         & IA527  & 26.4 & Subaru&7\\

         & IA598  & 26.2 & Subaru&7\\

         & V$_{606}\dagger$ & 26.6& {\it HST}&8\\

         & IA624  & 26.0 & Subaru&7\\

         & IA651  & 26.3 & Subaru&7\\

         & R         & 27.2 & VLT&1\\

         & IA679  & 26.2 & Subaru&7\\

         & IA738  & 26.1 & Subaru&7\\

         & IA767  & 25.1 & Subaru&7\\

         & z$_{850}\dagger$ & 25.6 & {\it HST}&8\\

         & Y           &24.5& VISTA&5\\

         & J            &24.7& CFHT&9\\

         & H           &23.8& VISTA&5\\

         & K           &24.1& CFHT&9\\
\hline
\hline
\end{tabular}
\end{center}
\end{table}

\section{Target selection}
The ideal situation when selecting targets for a spectroscopic survey
is to utilise a single photometric catalogue that provides consistent photometry with uniform wavelength coverage over the full survey area.
Unfortunately, this was not possible when performing target selection
for the VANDELS survey for two fundamental reasons. Firstly, given
that VANDELS targeted two separate survey fields,
covered by different sets of imaging data, it is clear that target
selection had to be performed using a minimum of two independent photometric catalogues.

Secondly, as described above, the footprint of the VANDELS survey
within the UDS and CDFS fields covers both the central areas with deep
{\it HST} imaging and the wider-field areas covered primarily by ground-based imaging (see Fig.~1). As a
result, the VANDELS survey area is effectively divided into four regions:
UDS-HST, UDS-GROUND, CDFS-HST and CDFS-GROUND, each of which required a separate photometric catalogue.
Consequently, the first stage in the target selection process was the adoption or production of robust photometric catalogues for each of
the four regions.

\subsection{Photometric catalogues}
Within the two regions covered by the WFC3/IR imaging provided by the
CANDELS survey (UDS-HST and CDFS-HST), we adopted the $H-$band selected photometric catalogues produced by the
CANDELS team (\citealt{gal13}; \citealt{guo13}). Both catalogues provide PSF-homogenised photometry for the available ACS and WFC3/IR
imaging, in addition to spatial-resolution matched photometry from
{\it Spitzer} IRAC and key ground-based imaging data sets derived using
the {\sc tfit} software package \citep{laidler07}. We refer the reader 
to \cite{gal13} and \cite{guo13} for full details of the production of
these photometric catalogues for the CANDELS UDS and CANDELS CDFS fields, respectively.

Within the wider-field areas there were
no publicly available, near-IR selected, photometric catalogues which met
our target selection requirements. As a result, new multi-wavelength
photometric catalogues were generated using the publicly available
imaging.  The imaging in both the UDS and CDFS fields was
initially accurately registered and placed on the same pixel scale and photometric zero-point. 
The imaging in the CDFS field had seeing which varied within the range
$0.6-1.0^{\prime\prime}$ FWHM. As a result, it was necessary to PSF-homogenise the images to a common spatial resolution of
$1.0^{\prime\prime}$ FWHM using Gaussian convolution kernels. The
imaging in the UDS field had a much narrower range of seeing ($0.8\pm 0.05^{\prime\prime}$~FWHM), meaning that
PSF-homogenisation was not necessary.

Following this initial processing, the photometric catalogues were
generated with {\sc sextractor} v2.8.6 \citep{sextractor} in dual-image
mode, using the $H-$band images as the detection images. Object
photometry was measured within
$2^{\prime\prime}-$diameter circular apertures, with accurate errors
calculated on an object-by-object basis using the
aperture-to-aperture variance between local blank-sky
apertures (see \citealt{mort17} for full details).

In Table 1 we provide details of the imaging data incorporated within the new photometric catalogues for the
UDS-GROUND and CDFS-GROUND regions. All of the depths listed in Table 1 refer to the data that were
publicly available and included in the target selection catalogues in summer 2015. 
We note that, since that date, many of the near-IR data sets have increased
in depth significantly, particularly within the extended CDFS field. 
 Therefore, to accompany the final data release of the VANDELS survey, we
are committed to publicly releasing updated photometric catalogues,
including deeper data where available, 
along with photometric redshifts and stellar-population parameters derived via SED fitting.

\subsection{Photometric redshifts}
\begin{figure}
\vspace{-0.5cm}
\begin{center}
\includegraphics[angle=0,width=0.39\textwidth]{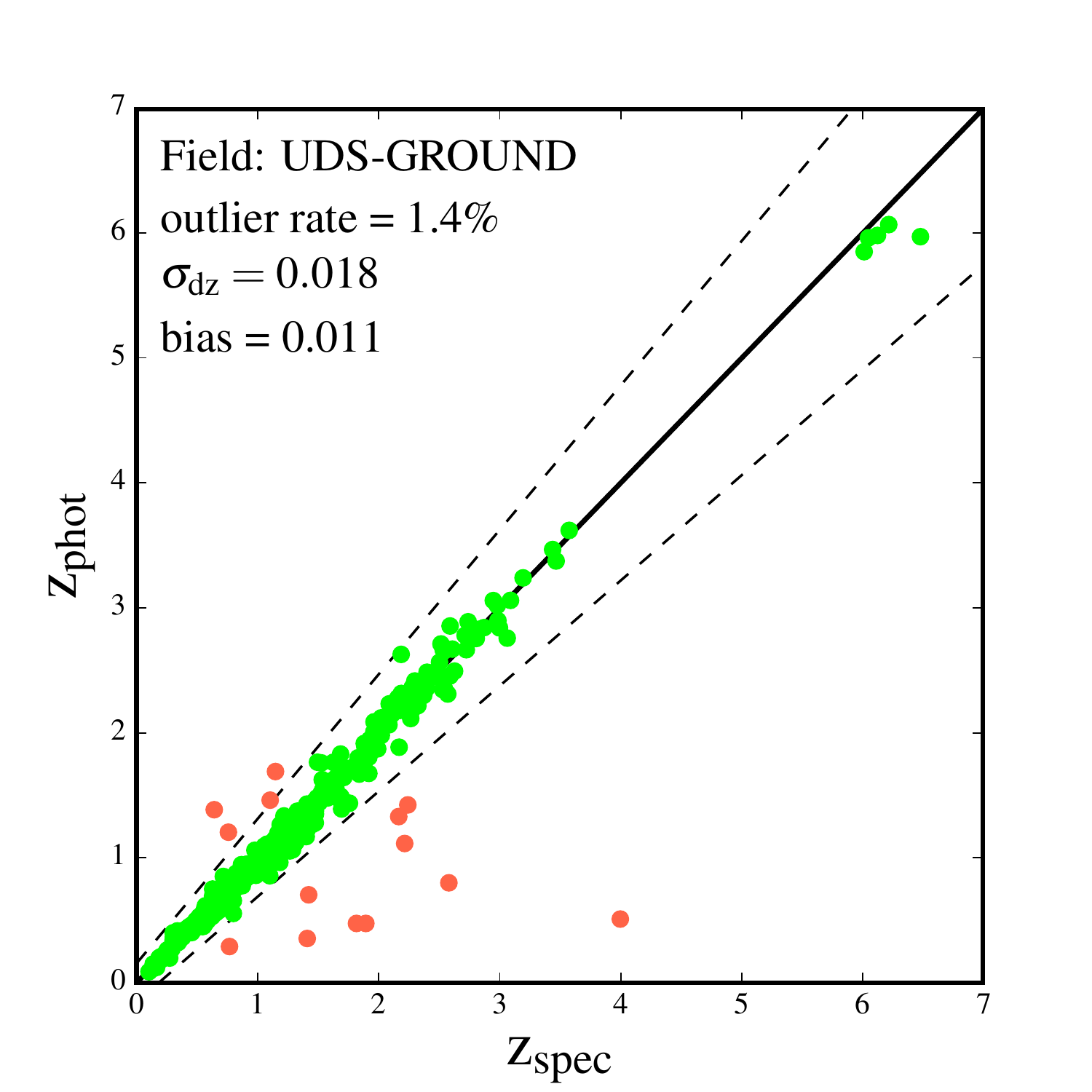}
\includegraphics[angle=0,width=0.39\textwidth]{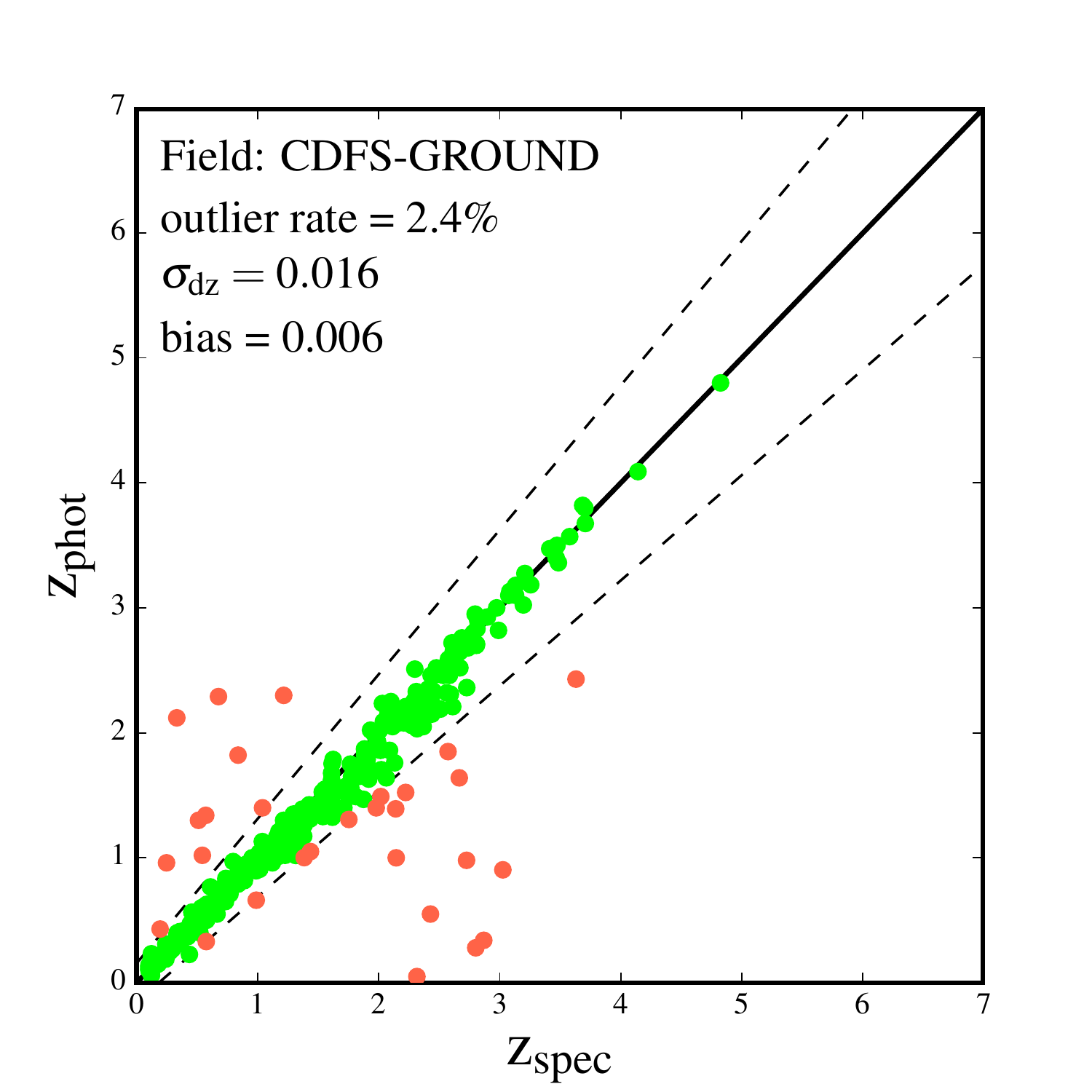}
\includegraphics[angle=0,width=0.39\textwidth]{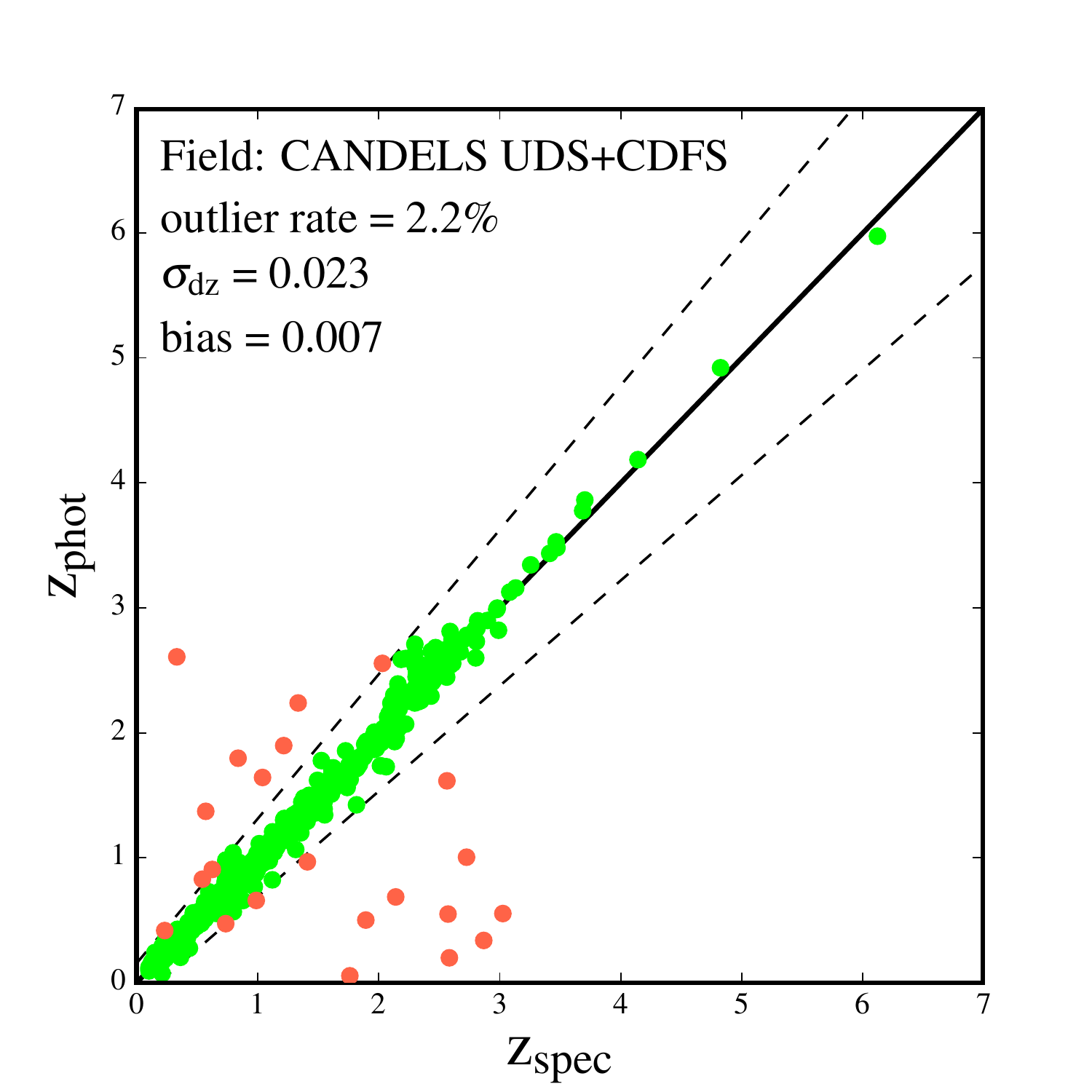}
\end{center}
\caption{Top: photometric redshifts derived by the VANDELS team
  compared to robust spectroscopic redshifts in the wide-area
  region of the UDS (red data-points are catastrophic outliers with
  $|{\rm dz}|>0.15$). Middle: equivalent plot for the wide-area
  region of the CDFS. Bottom: photometric redshift versus
  spectroscopic redshift for those objects in the top two panels for which photometric redshifts derived by the
  CANDELS survey team were available (see text for more details). The catastrophic outlier fraction, $\sigma_{\rm dz}$ and bias are
  displayed in the top-left corner of each panel.}
\end{figure}

A key element of the VANDELS survey strategy was the use of robust photometric-redshift pre-selection. 
For this process to be successful it was of paramount importance to 
either adopt or derive photometric redshifts of equal quality within all four of the VANDELS regions.
For the two regions covered by deep {\it HST} near-IR imaging
(UDS-HST and CDFS-HST), we adopted the photometric redshifts made
publicly available by the CANDELS survey team \citep{santini15}. As
discussed in \cite{dahlen13}, these photometric redshifts are derived
by optimally combining the independent estimates produced by a variety
of different photometric-redshift codes.

For the wider-area regions outside of the CANDELS WFC3/IR imaging
footprint, new photometric redshifts were generated within the VANDELS team, based
on the new UDS-GROUND and CDFS-GROUND photometric catalogues.  
These photometric redshifts were derived by taking the median value of
$z_{\rm phot}$ for each galaxy, based on a total of fourteen different
photometric redshift estimates derived by different members of the VANDELS team. 
The fourteen different photometric redshift estimates were produced
using a variety of different publicly-available codes
(e.g. \citealt{lephare1}; \citealt{hyperz}; \citealt{lephare2}; \citealt{eazy};
\citealt{zebra}) and in-house software (e.g. \citealt{fontana00}; \citealt{mclure11}), using a wide variety of different SED templates,
star-formation histories, metallicities and emission-line prescriptions.

In order to optimise their respective photometric-redshift codes, each member of the VANDELS team taking part in the
photometric-redshift exercise was initially allocated a spectroscopic training
set for the UDS-GROUND and CDFS-GROUND regions. Each training set consisted of approximately one thousand high-quality spectroscopic
redshifts, and were used by each team member to optimise the
performance of their code. The second step in the process was to allocate spectroscopic
validation sets to each member of the photometric-redshift team. The spectroscopic validation sets were identical in size and quality
to the corresponding training sets, the only difference being that the spectroscopic redshifts were not disclosed to the team members.
The accuracy of the results on these blind validation sets was used 
to ensure that each set of photometric-redshift estimates was adding
useful information to the overall result. For the UDS-GROUND region
the robust spectroscopic redshifts used for training and validation
purposes were drawn from the VIPERS survey
\citep{vipers}, the 3D-HST survey \citep{mom16} and the UDSz survey
(Almaini et al., in preparation). For the CDFS-GROUND region the
robust spectroscopic redshifts were drawn from the large number of
spectroscopic redshift campaigns previously undertaken  within the field
(e.g. \citealt{lefevre05}; \citealt{k20}; \citealt{vanzella08};
\citealt{balestra10}; \citealt{cooper12}; \citealt{lefevre13}; \citealt{mom16}).

To quantify the quality of the photometric redshift estimates we
calculate three statistics. To quantify any systematic off-set between
the photometric and spectroscopic redshifts we calculate the bias, which we define as the median
value of $dz~=~(z_{\rm spec}~-~z_{\rm phot})~/~(~1~+~z_{\rm spec})$. Secondly,
to quantify the accuracy of the photometric redshifts, we calculate
$\sigma_{\rm dz}$ using the robust median absolute deviation (MAD)
estimator. Finally, we also calculate the fraction of catastrophic
outliers, where an object is considered to be a catastrophic outlier
if $|dz|>0.15$. Based on the spectroscopic validation sets, the fourteen individual
photometric-redshift runs produced bias values in the range $0.03-0.003$, values of $\sigma_{\rm dz}$ in the range
$0.018-0.058$ and catastrophic outlier rates between 2\% and 16\%.
The equivalent statistics for the adopted median combined
$z_{\rm phot}$ results are bias= 0.008, $\sigma_{\rm dz}=0.017$ and a
catastrophic outlier rate of 1.9\%. Compared to the best-performing
individual photometric redshift run, the process of median combination
has produced a $15\%$ improvement in both $\sigma_{\rm dz}$ and
the catastrophic outlier fraction, with the same level of bias.
In Fig. 2 we show the accuracy of the final photometric redshifts adopted for the wider-area UDS-GROUND
and CDFS-GROUND regions, based on the spectroscopic validation sets.

Within the final spectroscopic validation sets used to define the
accuracy of the VANDELS photometric redshifts, 44\% of the galaxies also had photometric redshifts determined by the CANDELS team.
As a result, it was possible to perform a useful comparison of the quality of our
new photometric redshifts, based on the photometric data listed in Table 1, and the photometric redshifts derived by the CANDELS survey
team based on a combination of deep {\it HST} imaging, ground-based imaging and {\it Spitzer} IRAC imaging. 
For the objects in common, the VANDELS photometric redshifts have a catastrophic outlier rate of
2.0\% and $\sigma_{\rm dz}=0.018$, virtually identical to the
statistics for the full validation sets. The equivalent statistics for
the CANDELS photometric redshifts are an outlier rate of 2.2\% and
$\sigma_{\rm dz}=0.023$ (see bottom panel of Fig.~2).
The results of this comparison suggest that the VANDELS
photometric redshifts are slightly more accurate that the photometric redshifts derived by the CANDELS survey team. 

In summary, we are confident that by combining the results of the
CANDELS and VANDELS teams we were able to produce a final set of
photometric redshifts of consistent quality over all four of the VANDELS regions,
irrespective of the availability of deep {\it HST} imaging data. 

\subsection{Star-galaxy separation}
In order to produce the cleanest selection catalogue possible, it was
necessary to remove potential stellar sources. Due to the high angular
resolution provided by {\it HST}, this was a straightforward process
for the photometric catalogues within the UDS-HST and CDFS-HST regions. All
sources originating from the \cite{gal13} and \cite{guo13} catalogues
were excluded if they had a {\sc sextractor} \citep{sextractor} stellaricity
parameter of CLASS$\_$STAR $\geq 0.98$. Following the application of this criteria to remove stellar sources,
it was confirmed that the UDS-HST and CDFS-HST photometric catalogues
no longer displayed a stellar locus in a variety of different colour-colour diagrams.

For the two ground-based photometric catalogues, all sources consistent with the stellar locus on the $BzK$ diagram \citep{bzk} were excluded.
In addition, all remaining sources had their SED fitted with a range of
stellar templates drawn from the SpeX archive\footnote{http://pono.ucsd.edu/~adam/browndwarfs/spexprism/}.
All sources which produced an improved SED fit with a stellar template and were
consistent with being a point source at ground-based resolution were excluded. 
It should be noted that $<5\%$ of the objects in the two ground-based
photometric catalogues were excluded as being potentially stellar. 
Morevoer, it is noteworthy that $98\%$ of the excluded objects had $z_{\rm phot}<1$ and would therefore not even have
entered the VANDELS parent sample (see Section 4.5).

\subsection{Physical properties and rest-frame photometry}
At this stage, a final run of SED fitting was carried out in order to derive star-formation rates, stellar masses and rest-frame photometry. 
This SED fitting was performed using \cite{bc03} templates with solar
metallicity and no nebular emission. Exponentially-declining
star-formation histories were employed, with $\tau$ in the range
$0.3 \leq \tau \leq 20$ Gyr, and ages were constrained to lie between 50
Myr and the age of the Universe at the redshift of interest. Dust
attenuation was described using the \cite{calz00} starburst 
attenuation law, with $A_{\rm V}$ in the range $0.0 \leq A_{\rm V} \leq 2.5$, and IGM absorption was accounted for using the \cite{mad95} prescription.
These parameters were adopted following the results of \cite{wuyts11}, who showed that this parameter set does
a reasonable job of recovering the total star-formation rate of main-sequence galaxies, provided that they are not heavily obscured.
We also note that this SED parameter set is very similar to that
adopted by the 3D-HST survey team \citep{mom16} and delivers
stellar-mass estimates in good agreement with those derived for
the CANDELS CDFS and UDS photometric catalogues by \cite{santini15}.
During the SED-fitting process the redshift was fixed at the median value derived from the multiple photometric-redshift runs described in Section 4.2.

Further cleaning of the sample was carried out based on the results of
the SED fitting. For each of the four photometric catalogues, plots of the SED fits for the
objects comprising the worst 10\% of fits (i.e. highest $\chi^{2}$), were visually examined. 
Objects that were revealed by this process to have unreliable or
discrepant photometry were excluded from the sample ($\simeq 4\%$ of objects). 

\subsection{Parent spectroscopic sample}
Armed with catalogues providing robust photometry, photometric
redshifts and physical properties, it was then possible to select the parent
sample of {\it potential} spectroscopic targets. The vast majority (i.e. $\simeq 97$\%)
of the potential targets were drawn from three main target categories:
\begin{itemize}
\item{Bright star-forming galaxies in the range $2.4 \leq z \leq 5.5$}
\item{Lyman-break galaxies in the range $3.0 \leq z \leq 7.0$}
\item{Passive galaxies in the range $1.0 \leq z \leq 2.5$}
\end{itemize}
while the remaining $\simeq 3$\% of potential targets were either
known or candidate AGN ($\simeq 2$\%), or {\it Herschel-}detected galaxies ($\simeq 1$\%).

\subsubsection{Bright star-forming galaxies}
This sub-sample consists of bright star-forming galaxies within the
redshift range $2.4 \leq z \leq 5.5$
with $i\leq 25$. The redshift range is designed to ensure that the UV absorption features necessary for investigating stellar
metallicity lie within the $0.48 < \lambda < 1.0$~$\mu$m wavelength coverage of the VANDELS spectra. The magnitude constraint
is designed to ensure that the final VANDELS spectra have sufficient SNR to allow absorption-line studies on individual objects.
In order to be classified as actively star-forming, each member of this sub-sample was required to satisfy: sSFR > 0.1 Gyr$^{-1}$, where
sSFR is the specific star-formation rate (SFR/$M_{\ast}$) derived
from the SED fitting described in Section 4.4. In reality, 99\% of this sub-sample satisfy the criteria: sSFR > 0.6
Gyr$^{-1}$, ensuring that they are fully consistent with being located
on the main sequence of star formation (see Fig.~3).

\subsubsection{Lyman-break galaxies}
This sub-sample consists of fainter star-forming galaxies within the redshift
range $3.0 \leq z \leq 7.0$. The vast majority (95\%) of the galaxies
in this sub-sample lie in the redshift interval $3.0~\leq~z \leq~5.5$ and in the {\it HST} regions have $25 \leq H \leq 27
\land i \leq 27.5$. In the wider-field regions these objects have $i\leq 26.0$. 
The remainder of the sub-sample consists of galaxies selected to have redshifts in the range $5.5~\leq~z~\leq~7.0$ and, in the {\it HST} regions, to have $25 \leq H
\leq 27$ and $z^{\prime} \leq 26.5$ (UDS-HST) or $z_{850} \leq 27.0$ (CDFS-HST). In the wider-field regions these objects have $z^{\prime}\leq26.0$ and $z_{850}\leq25.0$ in
the UDS-GROUND and CDFS-GROUND regions, respectively. The change in selection
criteria for the $z \geq 5.5$ targets was mandatory, due to
the impact of IGM absorption on $i-$band photometry at these redshifts.
Once again, the formal requirement for these galaxies to be classified as star-forming was that sSFR > 0.1 Gyr$^{-1}$.
However, in reality, 99\% of the galaxies in this sub-sample have sSFR > 0.3
Gyr$^{-1}$ and provide a good sampling of the main sequence of star formation (see Fig.~3).

\subsubsection{Passive galaxies}
This sub-sample consists of $UVJ-$selected (\citealt{will09};
\citealt{whit11}) passive galaxies in the redshift interval $1.0\leq z
\leq 2.5$ with $H\leq 22.5 \land i\leq 25$. 
The $H-$band magnitude constraint for this sub-sample is designed to
impose an effective lower stellar-mass limit of $\log(M_{\ast}/\Msun)\geq 10$. As with the bright star-forming galaxy
sub-sample, the $i-$band magnitude constraint is designed to ensure
that the final individual spectra are deep enough to allow detailed
absorption-line studies. The $UVJ$ selection was performed using the rest-frame photometry derived from
the SED fitting described in Section 4.4. Galaxies which satisfied all
of the following criteria were identified as passive:
\begin{equation}
\begin{split}
U-V& > 0.88(V-J)+0.49,\\
U-V& > 1.2,\\
V-J &  <  1.6.
\end{split}
\end{equation}
We note here that although these galaxies are classified as passive,
it is not the case that they are necessarily expected to exhibit no
on-going star-formation. Based on the results of the SED fitting, 94\%
of the $UVJ-$selected passive galaxies do have estimated values of
sSFR<0.1 Gyr$^{-1}$, clearly separating them from main-sequence galaxies.
However, $3\%$ of the $UVJ-$selected passive galaxies have sSFR >
0.3 Gyr$^{-1}$, placing them in a location on the SFR$-M_{\ast}$ diagram consistent with the low-SFR tail of the main sequence.
This is not unexpected, given that $UVJ$ selection is inevitably vulnerable to contamination by dusty star-forming galaxies at some level.

\subsubsection{AGN and Herschel-detected galaxies}
The candidate AGN all lie within the CDFS field and were selected
based on either a power-law SED shape in the mid-IR
\citep{chang17} or X-ray emission (\citealt{xue11};
\citealt{rangel13}; \citealt{hsu14}). Within the CDFS-HST region the
candidate AGN were restricted to $z \geq 2.4$ and $i \leq 27.5$,
while in the CDFS-GROUND region they were restricted to $z \geq 2.4$ and $i \leq 26$. The {\it
  Herschel-}detected galaxies all lie within the UDS-HST and CDFS-HST
regions, have $z \geq 2.4$ and $i \leq 27.5$, and are detected in at least one {\it Herschel} band
(c.f. \citealt{pannella15}). We note here that the photometric
redshifts derived for the AGN candidates are based on SED fitting with
the same set of galaxy templates discussed in Section 4.2, and are
therefore not expected to be as accurate as the photometric redshifts
derived for the rest of the VANDELS sample.

\begin{figure*}
\begin{center}
\includegraphics[angle=0,width=0.9\textwidth]{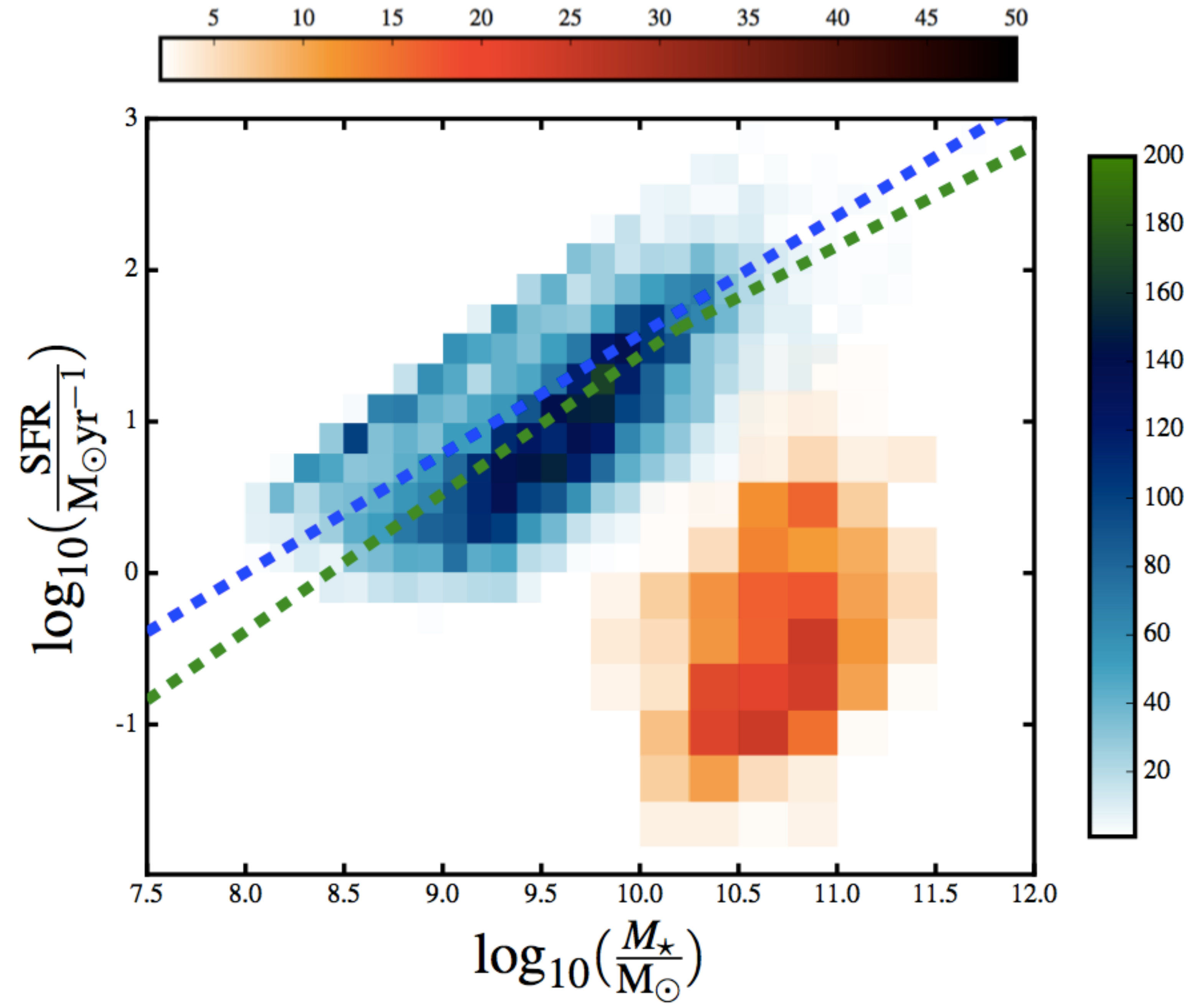}
\end{center}
\caption{The distribution of the VANDELS parent sample on the
  SFR$-M_{\ast}$ plane. The blue-shaded 2D histogram shows the
  location of the star-forming galaxies (including additional
  candidate AGN and
  {\it Herschel} sources) in the redshift interval $2.4 \leq z \leq 7.0$ (median
  redshift $z=3.4$). The red-shaded histogram shows the location of
  the passive galaxy sub-sample in the redshift interval $1.0 \leq z
  \leq 2.5$ (median redshift $z=1.2$). The horizontal and vertical colour bars
  indicate the number of galaxies within each 2D bin. The blue and green dashed lines show determinations of the main sequence of star-formation at
  $z=3$ and $z=2.5$ by \protect\cite{speagle14} and
  \protect\cite{whitaker14}, respectively. It can be seen that the VANDELS galaxies successfully sample the main
sequence of star-formation and the area of parameter space occupied by
massive, quenched galaxies. In total, the VANDELS spectroscopic sample 
spans 3.5 dex in stellar mass and 4.5 dex in star-formation rate.}
\end{figure*}

\subsubsection{Summary}
Following the application of the selection criteria outlined above, a
final visual check was performed on the entire sample to ensure that no image artefacts had
survived the selection procedure. The resulting parent sample of
potential VANDELS spectroscopic targets consisted of 9656 galaxies, split roughly equally between the UDS and CDFS fields.
The distribution of the parent sample on the SFR$-M_{\ast}$ plane is
shown in Fig. 3, from which it can be seen that the adopted selection
criteria successfully isolated the main sequence of star formation
and the high stellar-mass quenched population. Overall, the parent VANDELS sample spans 3.5 dex in stellar mass and 4.5
dex in star-formation rate.

\subsection{Final spectroscopic sample}
\begin{table}
\caption{The distribution of the 2106 spectroscopic slits 
targeted within the VANDELS survey between the two survey fields, the
  different target classifications and the different integration
  times. The first column lists the survey field. Column two lists the number
  of slits allocated to bright star-forming galaxies (SFG), column
  three lists the number of slits allocated to massive, passive
  galaxies (PASS), column four lists the number of slits allocated to
  fainter star-forming galaxies (LBG) and the fifth column lists the
  number of slits allocated to AGN candidates or {\it
    Herschel-}detected galaxies (AH).  Note that all of the AGN
  candidates were selected in the CDFS field due to the availability
  of ultra-deep X-ray data \protect\citep{luo17}.The final three columns list
  the number of slits allocated to objects which require 20, 40 and 80
  hours of on-source integration, respectively.}
\begin{tabular}{lccccccc}
\hline
FIELD& SFG & PASS & LBG & AH & 20 & 40 & 80\\
\hline
\hline
UDS   & 224 & 151   & 693 & 10       &303 & 550 & 225\\
\hline
CDFS & 200 & 117 & 656 & 55 & 238 &528 & 262 \\
\hline
TOTAL & 424 & 268 & 1349 & 65 &541 &1078 & 487\\
\hline
\hline
\end{tabular}
\end{table}

\begin{figure}
\begin{center}
\includegraphics[angle=0,width=0.5\textwidth]{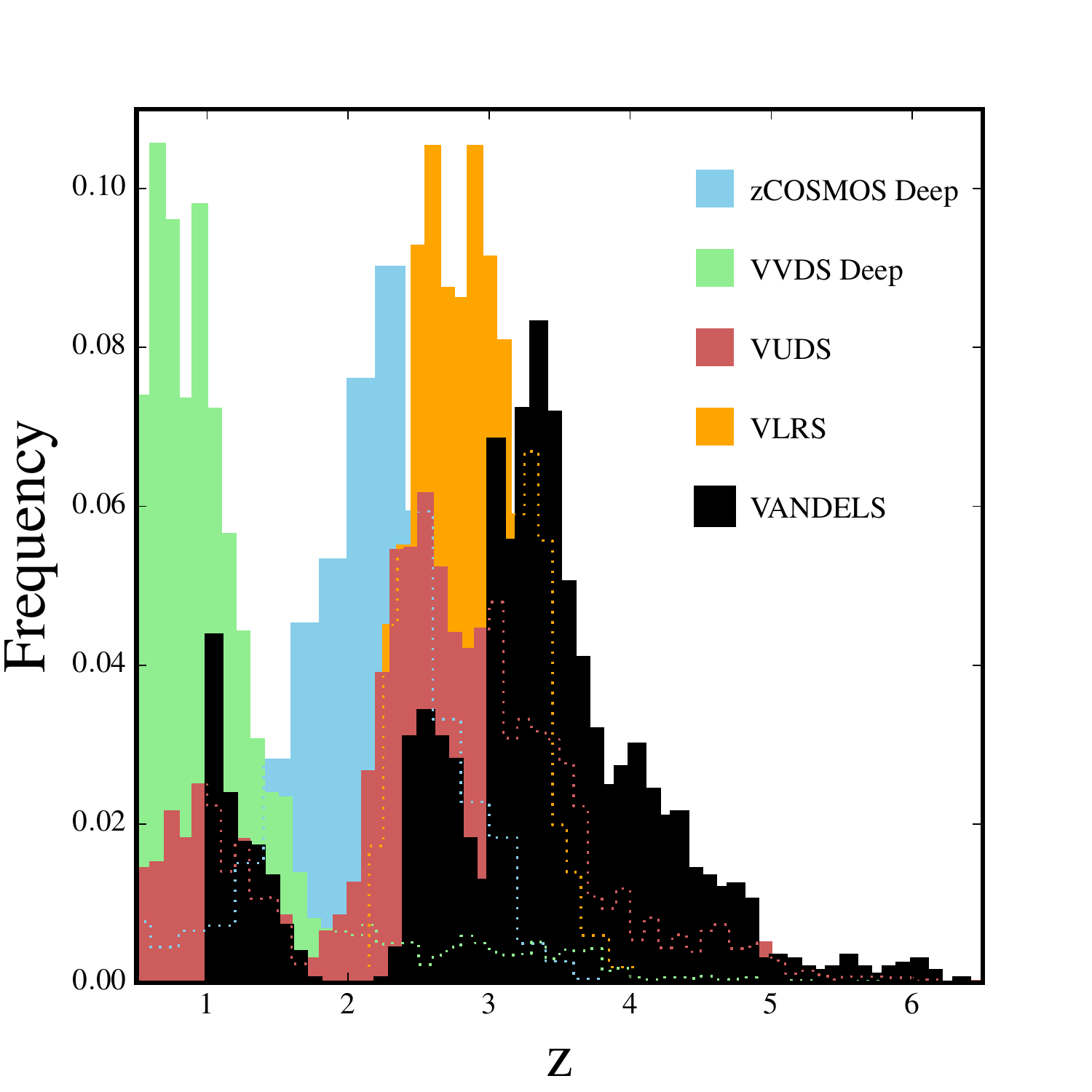}
\end{center}
\caption{A comparison of the redshift distributions of large-scale
  spectroscopic surveys carried out with the VIMOS spectrograph. The
  deep component of the zCOSMOS survey \citep{lilly07} is shown in
  blue and the deep component of the VIMOS VLT Deep Survey (VVDS) is shown in green
  \citep{lefevre13}. The VIMOS Ultra Deep
  Survey (VUDS) is shown in red \citep{lefevre15} and the VLT LBG Redshift Survey (VLRS) is
  shown in orange \citep{bielby13}. The black histogram shows the photometric
  redshift distribution of the final sample of 2106 galaxies targeted by the
  VANDELS survey.}
\end{figure}
Using the parent sample as input, extensive simulation work was 
undertaken in order to maximise the number of slits which could be allocated across the eight VIMOS pointings.
In addition to the total number of spectroscopic slits, the primary 
goal of this experimentation was to maximise the number of slits
allocated to bright star-forming galaxies and massive passive
galaxies, the two classes of targets with the lowest surface
densities. Apart from the photometric redshift and magnitude
constraints outlined above, the only additional constraint applied to
the simulations was the desire to allocate the slits to objects
requiring 20, 40 and 80 hours of integration in an approximately 1:2:1 ratio.
Crucially, during the slit allocation process, no additional
prioritisation was applied based on source brightness, redshift or position.

The overall result of the target selection process was a final sample of 2106
galaxies being allocated to spectroscopic slits. The distribution of the spectroscopic slits
between the two survey fields, the different target classifications
and the different amounts of required exposure time are detailed in Table 2.
The final spectroscopic samples of bright star-forming
galaxies and passive galaxies are random (approximately 1 in 4)
sub-samples drawn from the corresponding targets within the input
parent spectroscopic sample. Likewise, the final spectroscopic sample
of Lyman-break galaxies is a random
(approximately 1 in 5) sub-sample of the Lyman-break targets within the parent spectroscopic sample.
In Fig.~4 we compare the photometric-redshift distribution of the
final VANDELS sample to the spectroscopic redshift distributions of 
comparable large-scale spectroscopic surveys previously carried out using the VIMOS spectrograph.

\section{Observing strategy}
As illustrated in Fig.~1, the VANDELS survey consists of a total of eight VIMOS pointings, four
overlapping pointings in UDS and four overlapping pointings in
CDFS. In both fields the pointing centres were chosen to provide both contiguous
coverage and to fully sample the central areas with deep {\it HST}
imaging. Fully covering the deep {\it HST} imaging was essential in order 
to allow access to a high surface-density of faint $z\geq 3$ targets.

\subsection{Signal-to-noise requirements}
The VANDELS observing strategy was designed to provide consistently
high SNR continuum detections for the bright star-forming and passive
galaxy sub-samples. For those objects with $i\leq 24.5$, the final 1D spectra are designed
to have a SNR in the range $15-20$ per resolution element, within the wavelength range $6000 < \lambda < 7400$~\AA,
based on 20 or 40 hours of on-source integration (where one resolution
element is 4 pixels, or 10.2\AA). For the faintest objects in these sub-samples ($i\simeq25$), the final
spectra are designed to have SNR $\simeq 10$, based on 80 hours of integration.
For the fainter ($H\leq27 \land i\leq27.5$) Lyman-break galaxies at $z\geq3$,
the VANDELS observing strategy is designed to provide SNR $\geq 3$
in the continuum, and a  consistent Ly$\alpha$
emission-line detection limit of $\simeq 2\times10^{-18}$
erg s$^{-1}$cm$^{-2}$ ($5\sigma$, integrated over a line profile with FWHM=10\AA).

In order to achieve the desired SNR, targets were allocated 20, 40 or
80 hours of on-source integration according to two different exposure
time schemes. The bright star-forming and passive galaxies were allocated 20 hours of integration time if $i_{2}\leq 23.75$, 40
hours in the range $23.75< i_{2} \leq 24.25$ and 80 hours in the range
$24.25 < i_{2} \leq 25.00$ (where $i_{2}$ is
the $i-$band magnitude measured in a 2$^{\prime\prime}-$diameter
circular aperture at ground-based resolution\footnote{the typical off-set
  between $i_{2}$ and the total $i-$band magnitudes used throughout
  the rest of the paper is $\simeq 0.3$ mag.}).
The LBGs, AGN candidates and {\it Herschel-}detected galaxies were allocated 20, 40
or 80 hours of integration time within the following three magnitude ranges: $25.00< i_{2}
\leq 25.50$, $25.50< i_{2} \leq 26.00$ and $26.00< i_{2} \leq
27.50$. The highest-redshift LBG targets at $z\geq 5.5$ followed the
same exposure time scheme as the main LBG sub-sample, except with the $i-$band magnitudes replaced with $z-$band magnitudes. 

\subsection{Nested slit allocation policy}
To accommodate the required range of exposure times, the VANDELS survey employed
a nested slit allocation strategy. Each of the eight VIMOS pointings was observed using four sets of masks, with each set
receiving 20 hours of on-source integration time. Consequently, objects which
required 80 hours of integration were retained on all four masks, those
requiring 40 hours were included on two masks and those requiring 20 hours only appeared on a single mask.
As can be seen from Table 2, approximately 75\% of the galaxies
targeted by the VANDELS survey received 40+ hours of on-source integration.

\subsection{Observations}
All of the VANDELS observations used the MR grism+GG475 order sorting
filter, 1 arcsec slit widths and a minimum slit length of 7 arcsec.
This set-up provides wavelength coverage of 480$-$1000 nm, with a dispersion of 0.255 nm/pix and a mean spectral resolution of $R\simeq 580$. 
All of the slits were oriented E-W on the sky, as recommended for minimising
slit losses when pursuing long integrations of the UDS and CDFS fields
from Paranal \citep{sanchez}. To ensure that the VIMOS slits were placed with maximum accuracy,
short $R-$band pre-images were obtained in service mode during P94, in
order to properly account for VIMOS focal plane distortions and allocate 1--2 bright
reference stars to each VIMOS mask.

All observations were obtained using observing blocks (OBs) designed to deliver a total
of one hour of on-source integration time. Each OB consisted of three
integrations of 1200s, obtained in a three-point dither pattern, with
off-sets of 0, $-$4 pixels, +8 pixels, corresponding to 0.0, $-0.82$ and
+1.64 arcsec respectively. One arc frame and one flat-field frame
were obtained for calibration purposes after the execution of  two consecutive
OBs. A spectrophotometric standard was observed at least once every
seven nights and at least once per observing run. Further details of
the VANDELS observations can be found in the data release
paper \citep{dr1paper}.

\section{Data Reduction and Spectroscopic Redshift Measurement}
The reduction of the VANDELS data set is performed with the fully-automated {\sc easylife} pipeline, 
starting from the raw data and ending with the fully wavelength- and flux-calibrated one-dimensional spectra. 
The {\sc easylife} pipeline \citep{garilli12} is an updated version of the original 
{\sc vipgi} system \citep{scod05}. The original
{\sc vipgi} system was used to reduce all the spectra from the VVDS (\citealt{lefevre05}; \citealt{garilli08}), zCosmos \citep{lilly07} and 
VUDS surveys \citep{lefevre15}, while the updated system {\sc easylife} was used to reduce all of the spectra from the
recently completed VIPERS survey \citep{guzzo14}.  A detailed description of the full data reduction process can be found in \cite{dr1paper}.

In addition to the reduced spectra, it is a requirement of the ESO
public survey agreement for VANDELS that the team provide
spectroscopic redshift measurements for each of the spectra released via the ESO data archive.
The spectroscopic redshift measurements were made by a dedicated group
of VANDELS team members using the {\sc ez} software package
\citep{garilli10}. The core algorithm of {\sc ez} is cross-correlation using galaxy templates that, for VANDELS spectra,
were predominantly derived from previous VIMOS surveys. The redshift for each galaxy was independently measured by two team
members, who were subsequently required to reach agreement on the
spectroscopic redshift measurement and the associated quality flag. 
As a final check, the spectroscopic redshifts and associated quality flags for all spectra released in DR1 were independently checked by the two Co-PIs.

The quality of the spectroscopic redshift measurements was quantified
using the system originally employed by the VVDS team (\citealt{lefevre05}),
in which every galaxy is allocated a quality flag of 0, 1, 2, 3, 4 or 9. 
Galaxies for which it was not possible to measure a spectroscopic
redshift are allocated flag=0, while galaxies with spectroscopic
redshift measurements that are believed to be 50\% or 75\% reliable
are allocated flag=1 and flag=2, respectively. The galaxies with the most secure
redshifts, based on multiple absorption/emission features, are
allocated flag=3 or 4, depending on whether their redshift
measurements are believed to be 95\% or 100\% reliable.
Galaxies which have redshift measurements based on a single emission
line, in most cases Ly$\alpha$, are allocated flag=9.

\begin{figure*}
\begin{center}
\includegraphics[angle=0,width=0.7\textwidth]{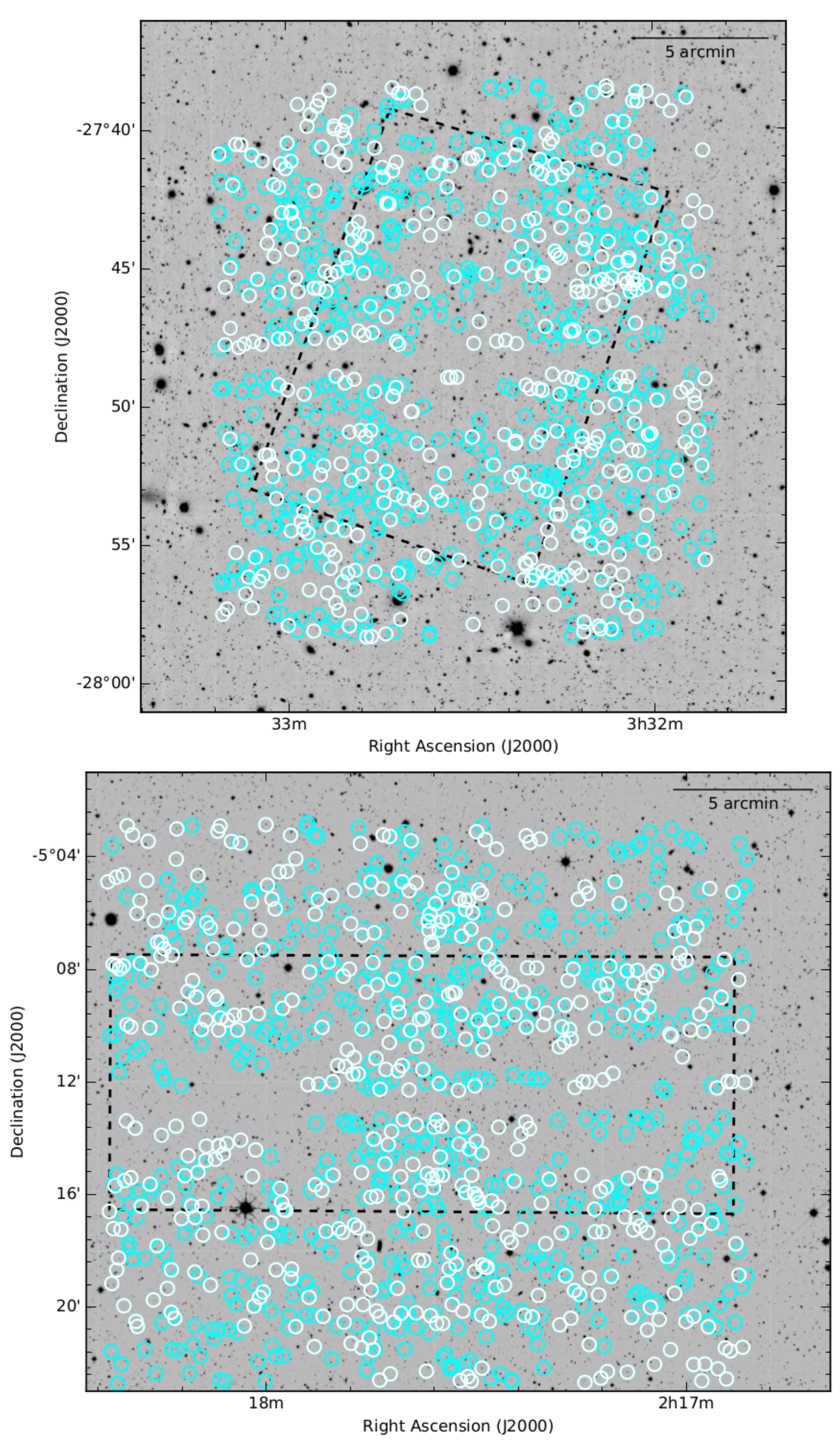}
\caption{Finding charts showing the location of the VANDELS
  spectroscopic targets within the CDFS (top) and UDS (bottom) fields. 
The 415 targets in the CDFS and 464 targets in the
  UDS with spectra released in VANDELS DR1 are shown in white, with the
  remaining targets shown in blue. The black dashed rectangles show the approximate location of
  the CANDELS near-IR {\it HST} imaging (\citealt{grogin11}; \citealt{koke11}). The background images are ground-based $H-$band data from
  the VISTA VIDEO (\protect\citealt{jarvis13})  and UKIDSS UDS
  (Almaini et al., in preparation) surveys.}
\end{center}
\end{figure*}

\begin{figure*}
\includegraphics[angle=0,width=0.8\textwidth]{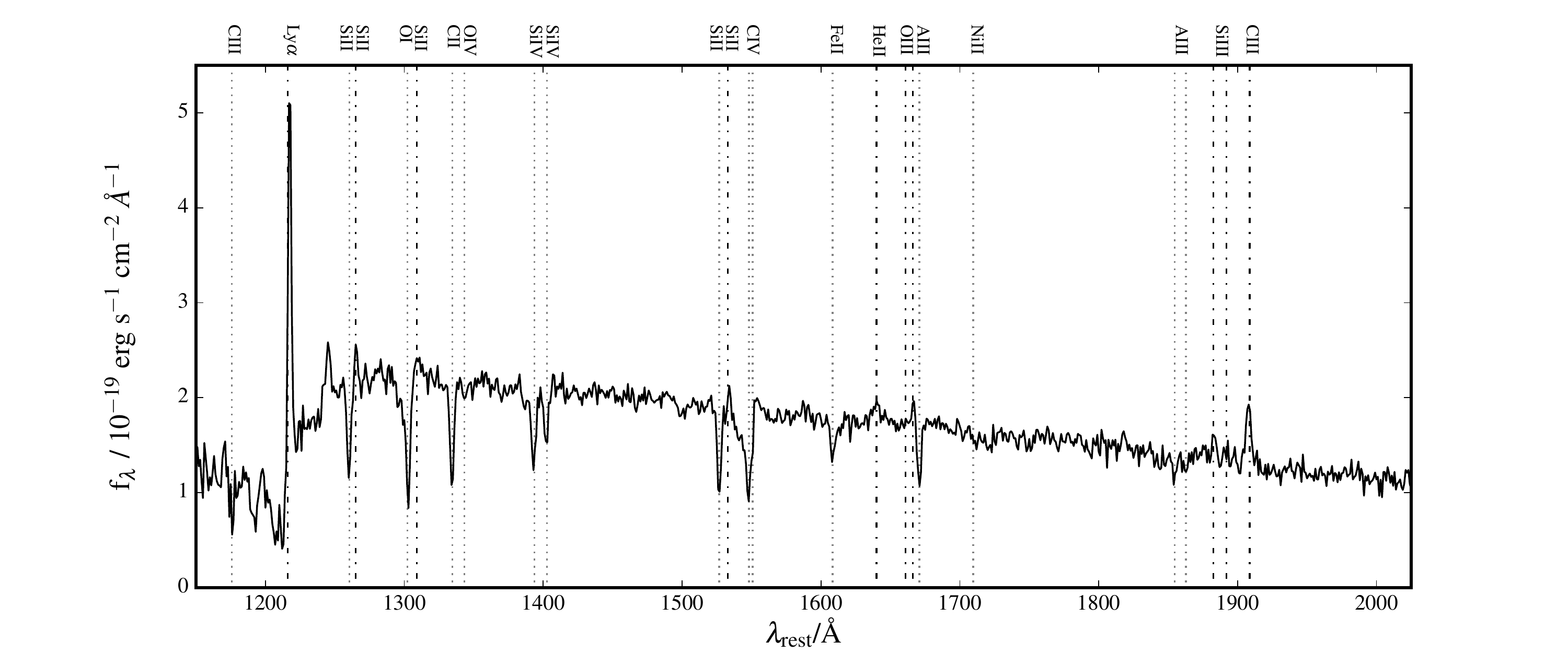}
\includegraphics[angle=0,width=0.8\textwidth]{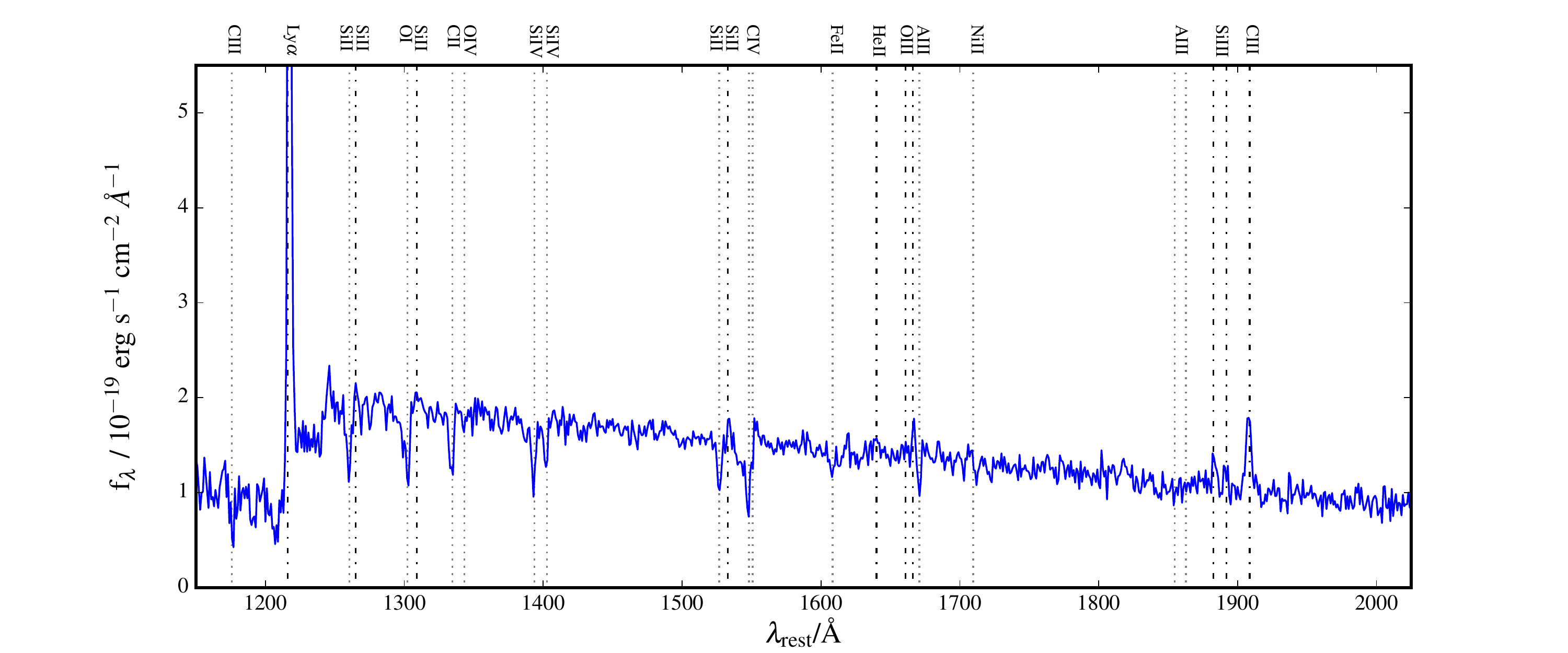}
\includegraphics[angle=0,width=0.8\textwidth]{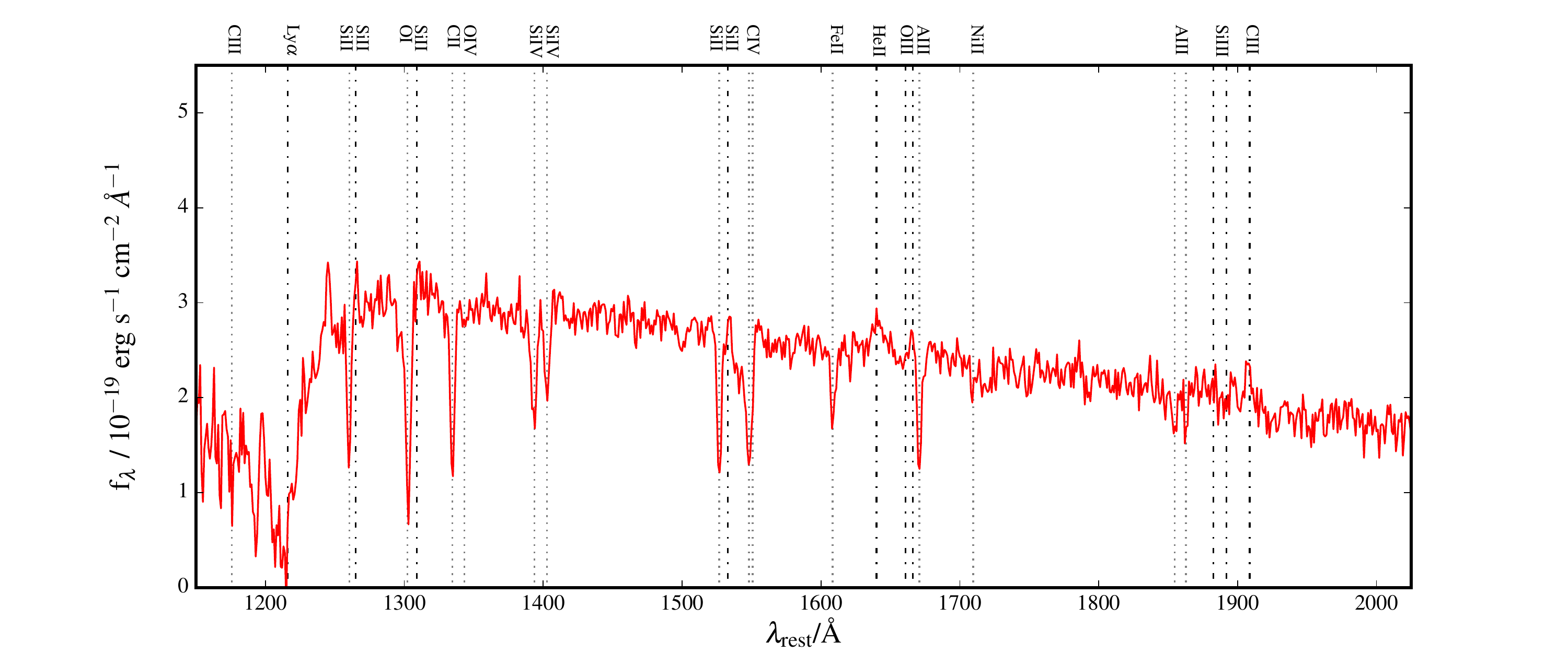}
\caption{Median-stacked spectra of Lyman-break galaxies from VANDELS DR1. The
  top panel shows a stack of 105 LBGs from
  DR1 with robust redshifts in the range $3.0 \leq z \leq 4.0$ (median redshift
  $z=3.5$). The middle panel shows a stack of the
  61/105 galaxies that display Ly$\alpha$ in
emission. The bottom panel shows a stack of the 44/105 galaxies
that display Ly$\alpha$ in absorption. In
all three panels, common absorption (dotted lines) and emission (dot-dashed lines)
features are highlighted.}
\end{figure*}

\begin{figure*}
\includegraphics[angle=0,width=0.8\textwidth]{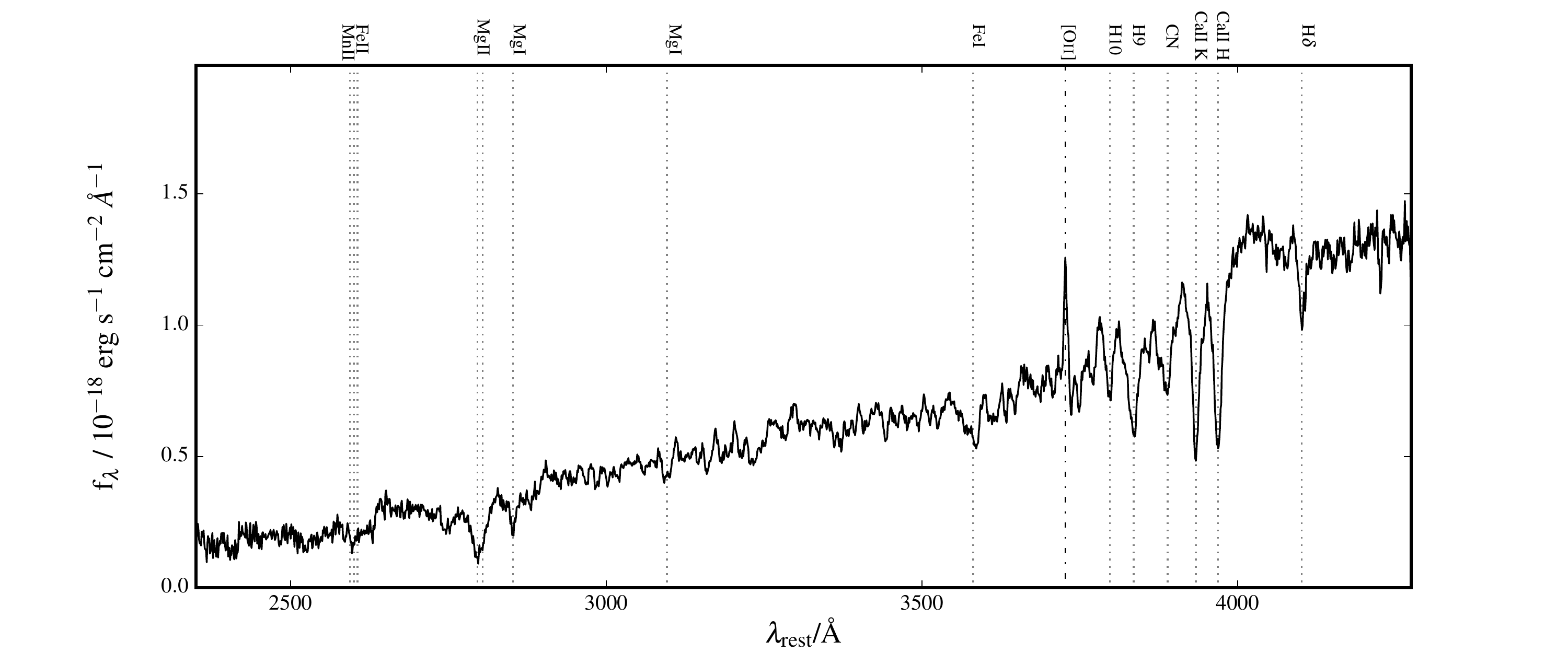}
\includegraphics[angle=0,width=0.8\textwidth]{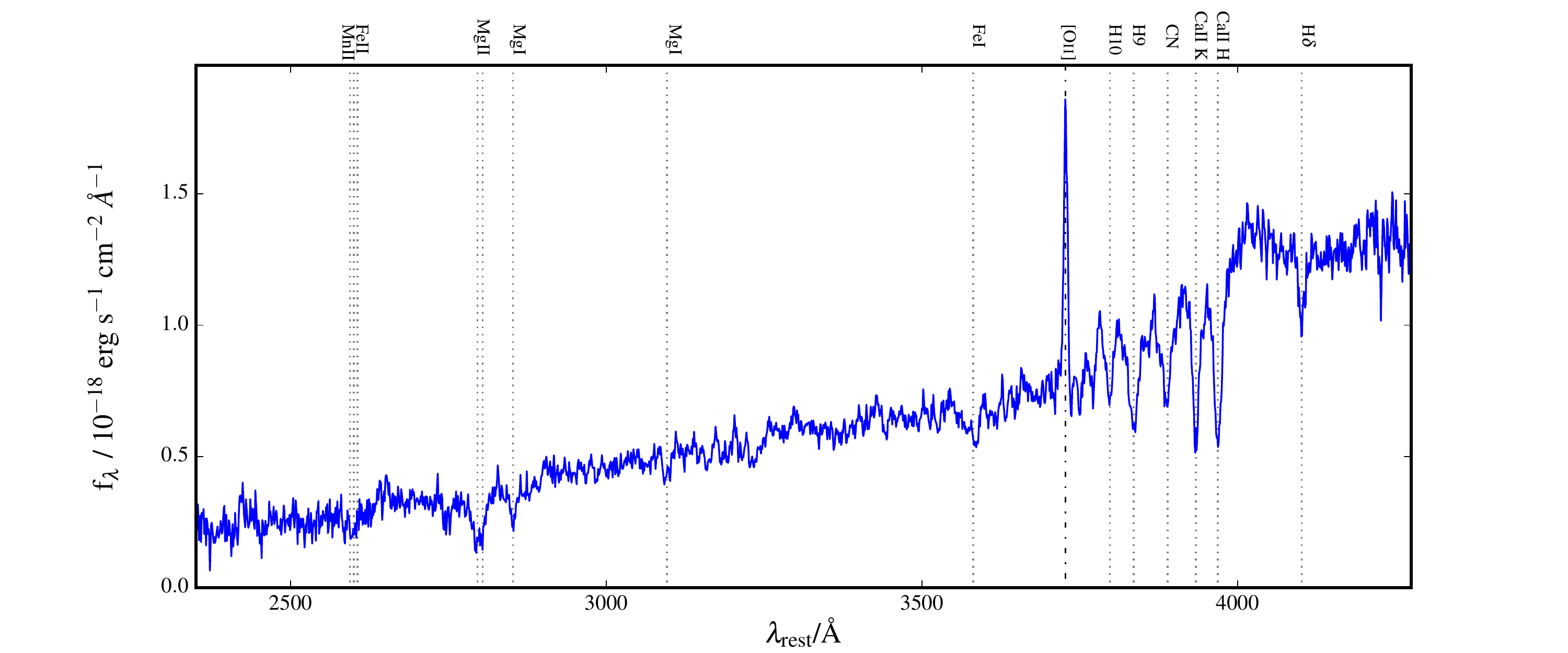}
\includegraphics[angle=0,width=0.8\textwidth]{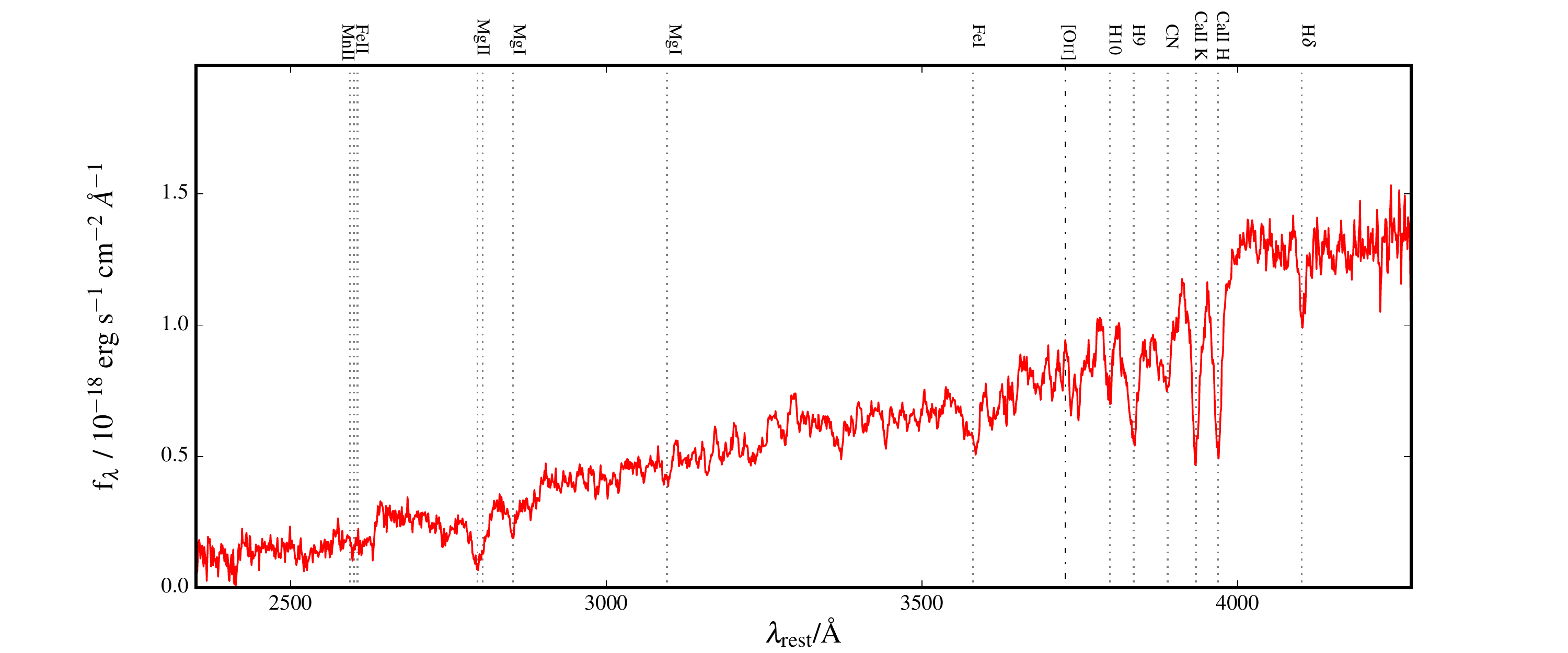}
\caption{Median-stacked spectra of passive galaxies from VANDELS DR1. The top panel shows a stack of 65 passive galaxies from
  DR1 with robust redshifts in the range $1.0 \leq z \leq 2.5$ (median redshift $z=1.2$). The middle panel shows a stack of the
  33/65 passive galaxies that display [O{\sc ii}] emission. The bottom panel shows a stack of the 32/65 passive galaxies without [O{\sc ii}] emission.
Common absorption (dotted lines) and emission (dot-dashed lines) features are highlighted in each panel.}
\end{figure*}

\begin{figure}
\vspace{-0.5cm}
\begin{center}
\includegraphics[angle=0,width=0.38\textwidth]{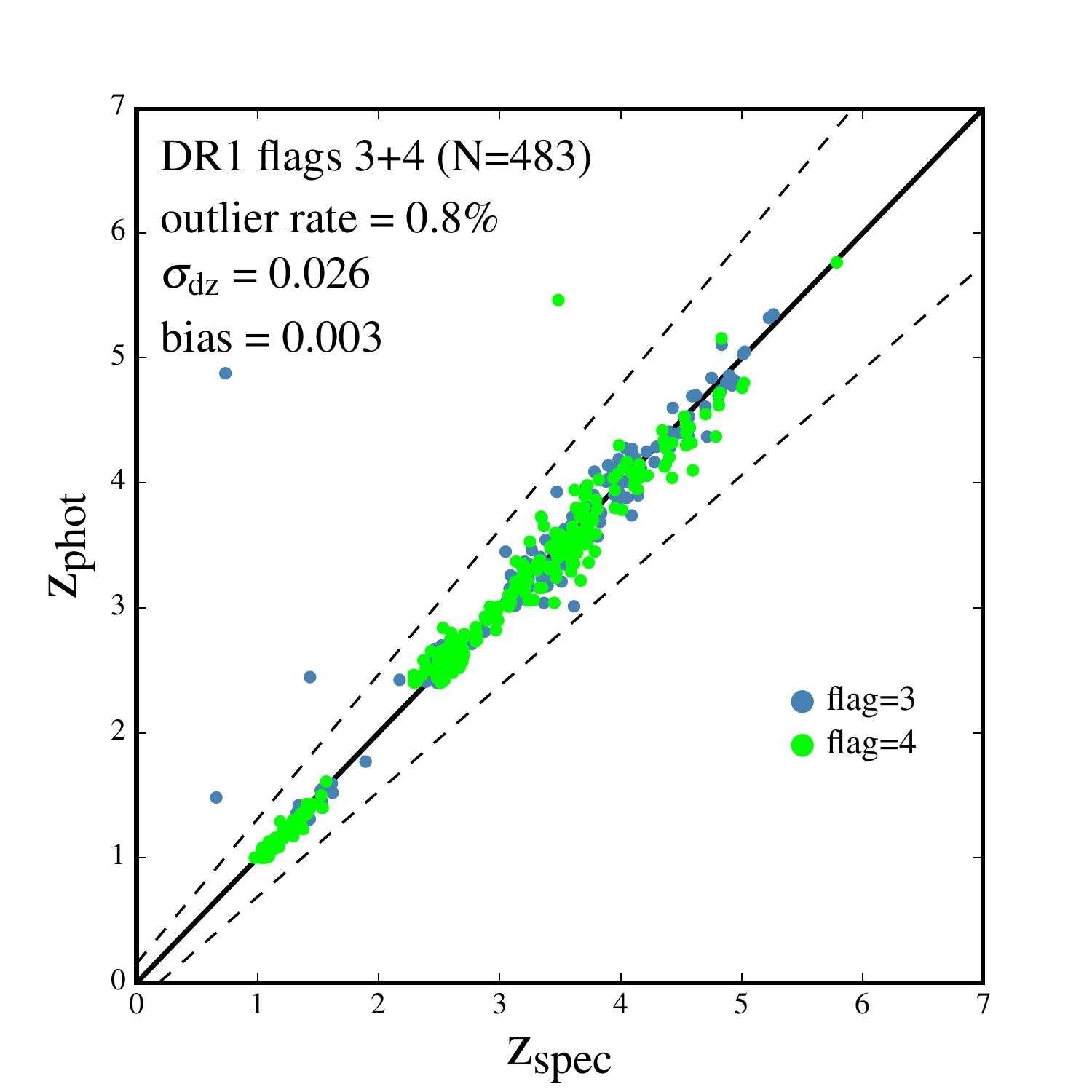}
\includegraphics[angle=0,width=0.38\textwidth]{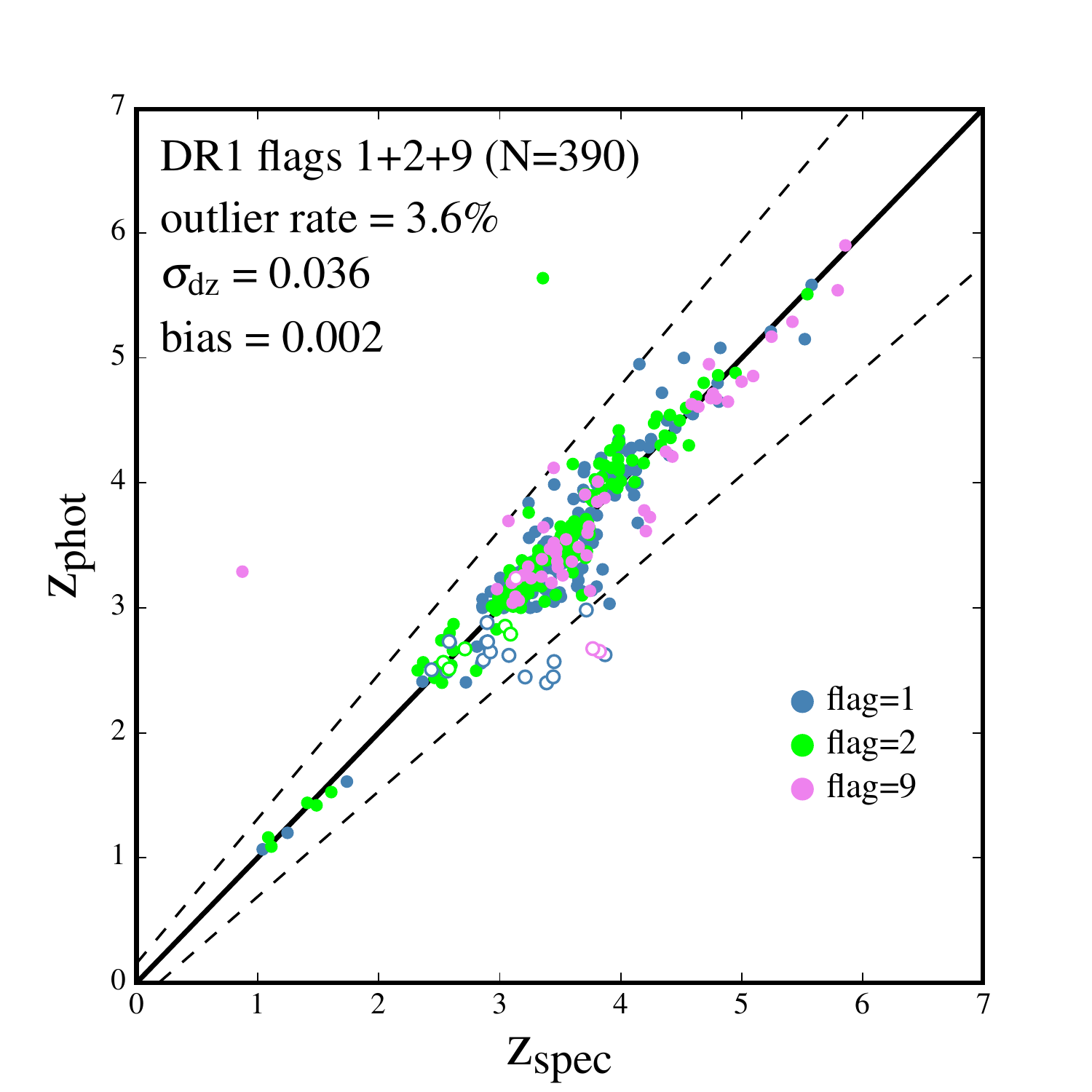}
\includegraphics[angle=0,width=0.38\textwidth]{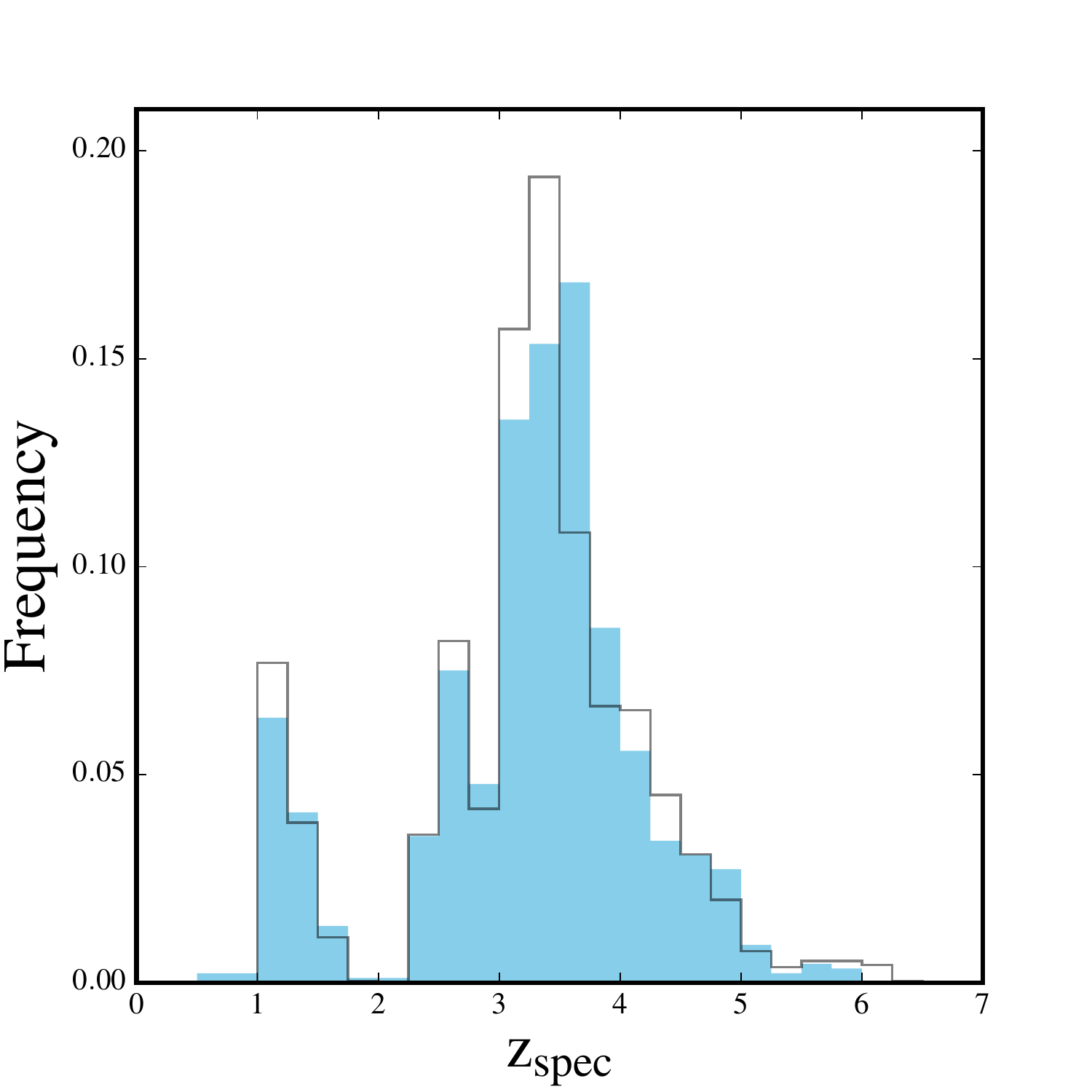}
\end{center}
\vspace{-0.2cm}
\caption{The top panel shows a comparison between the input photometric
  redshifts and measured spectroscopic redshifts for DR1 galaxies with
  redshift quality flags 3 and 4. The middle panel is the
  equivalent plot for DR1 galaxies with redshift quality flags 1, 2
  and 9. Those galaxies falling outside the dashed lines are catastrophic
  outliers with $|{\rm dz}|>0.15$. In both panels, candidate AGN and {\it
    Herschel-}detected galaxies are plotted as open symbols.
The bottom panel shows a comparison of the spectroscopic redshift distribution of
  the DR1 galaxies (solid blue histogram) and the photometric redshift
  distribution of the full VANDELS parent sample (open histogram).}
\end{figure}

\begin{figure}
\vspace{-0.5cm}
\begin{center}
\includegraphics[angle=0,width=0.39\textwidth]{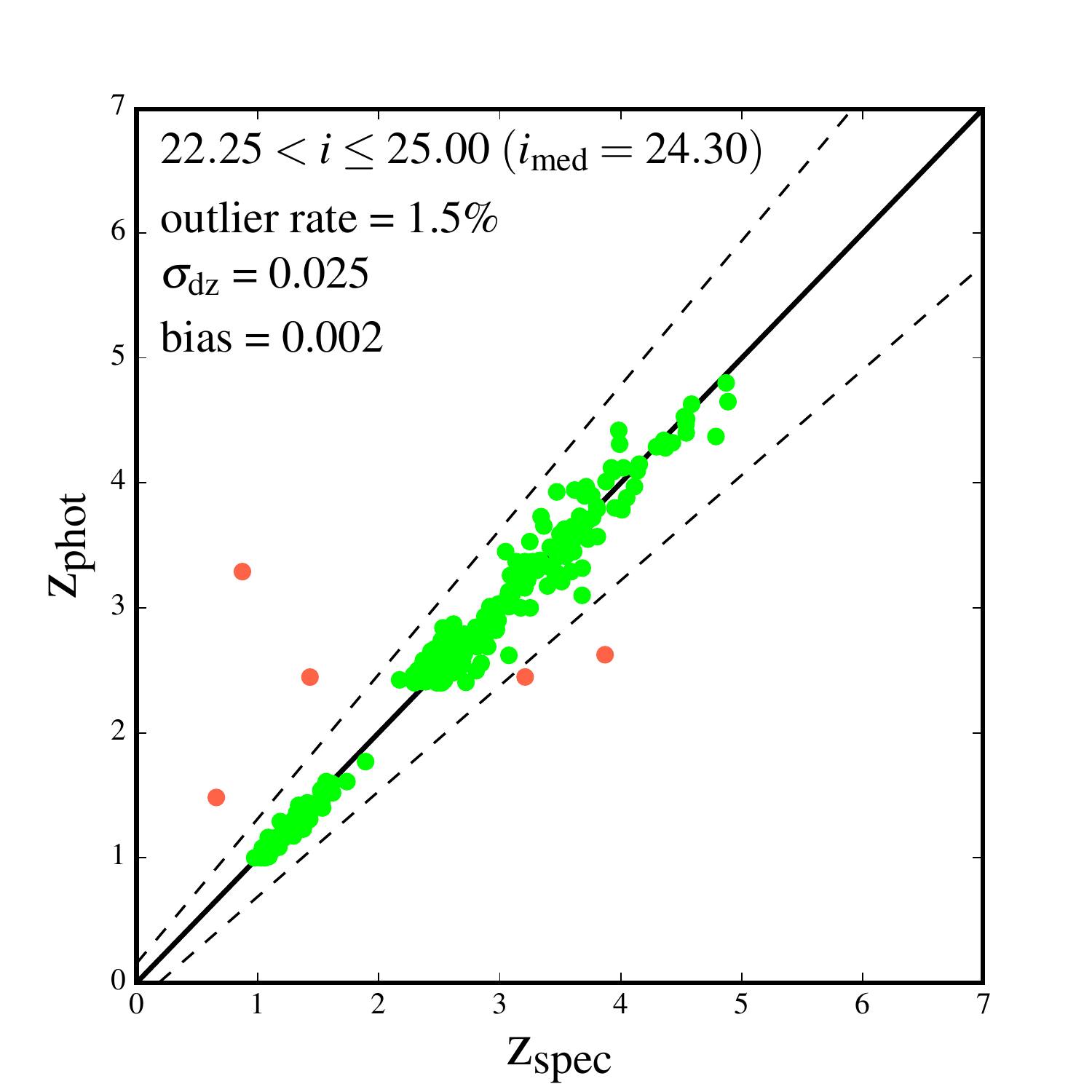}
\includegraphics[angle=0,width=0.39\textwidth]{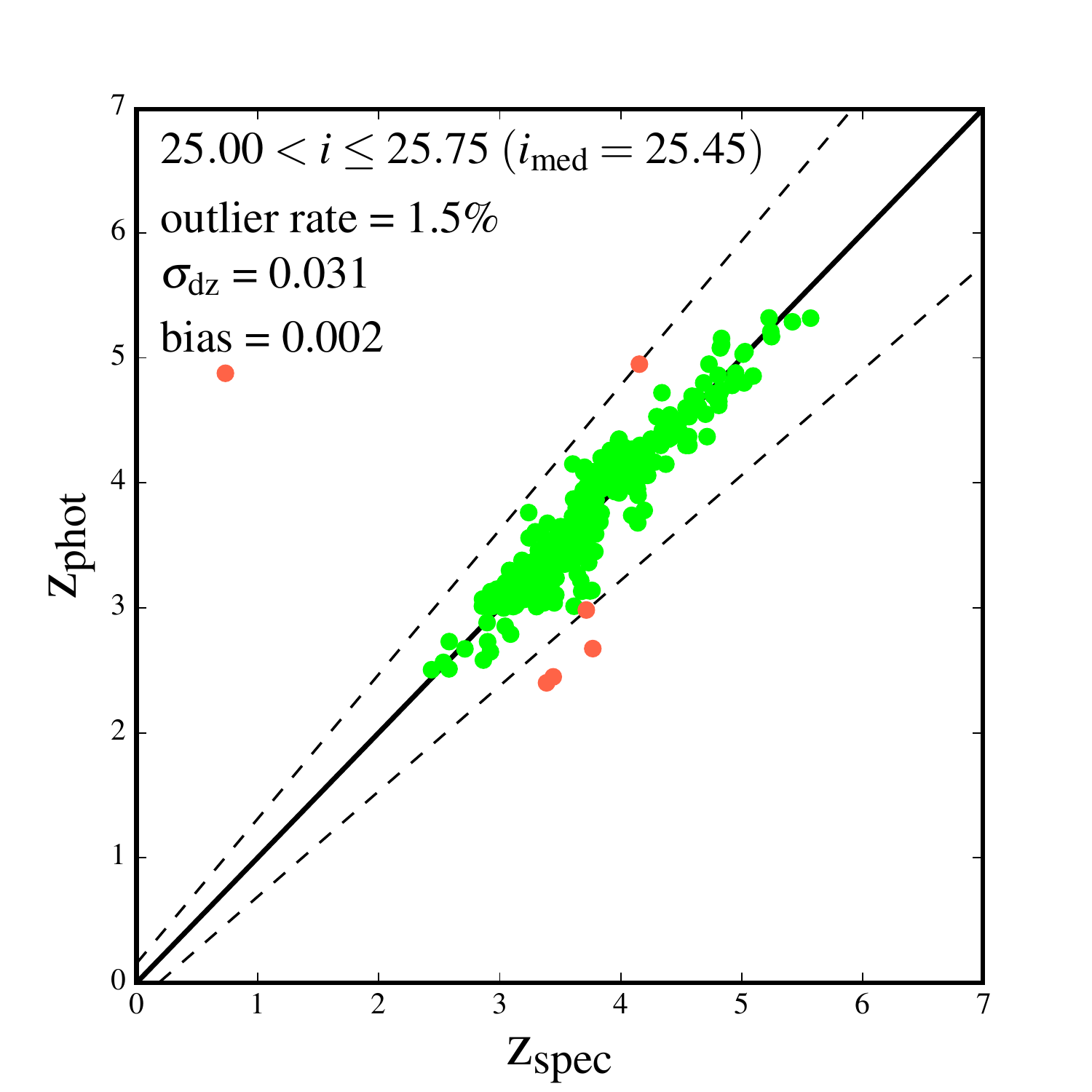}
\includegraphics[angle=0,width=0.39\textwidth]{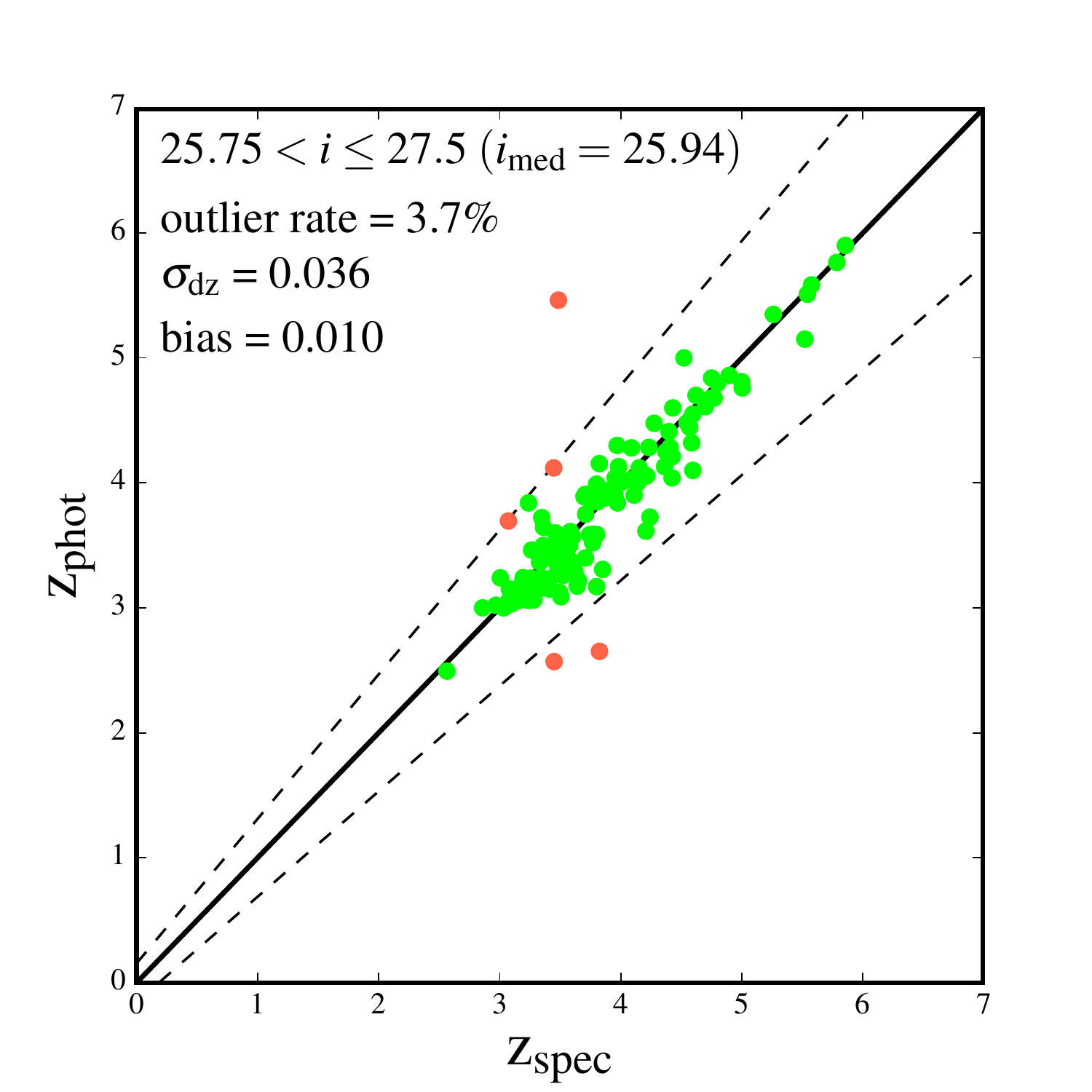}
\end{center}
\caption{The top panel shows a comparison between the input photometric
  redshifts and measured spectroscopic redshifts for DR1 galaxies in
  the magnitude range $22.25 < i \leq 25.00$. The middle and bottom
  panels show the equivalent plots for DR1 galaxies in the magnitude
  ranges $25.00 < i \leq 25.75$ and $25.75 < i \leq 27.50$,
  respectively. All three panels include all DR1 galaxies with
  spectroscopic redshift quality flags in the range $1-9$.}
\end{figure}

\section{Data Release One}
The first public data release for the VANDELS survey (DR1) was made by the ESO Science Archive Facility
({\tt archive.eso.org}) on 29th September 2017, and features spectra
obtained during the first VANDELS observing season from August 2015 until February 2016; ESO run numbers 194.A-2003(E-K).
The data release includes fully flux- and wavelength-calibrated 1D
spectra, plus wavelength calibrated 2D spectra, for all the VANDELS
targets that received their total scheduled integration time during season one. In
addition, the data release also includes spectra for those targets that had received 50\% of their scheduled integration time by the end of season one.

In total, DR1 contains spectra for 879 galaxies, 415 from the CDFS pointings and 464 from the UDS pointings. In Fig. 5 we show finding
charts for the CDFS and UDS fields which show the locations of the full VANDELS target list in blue,  with the locations of those VANDELS
targets featured in DR1 in white. In addition to the reduced spectra, DR1 also features an associated
catalogue which provides coordinates, optical+nearIR photometry, photometric redshifts, spectroscopic redshifts and
spectroscopic redshift quality flags for each target. In Figs.~6 \& 7, we show examples that illustrate the potential for
using the DR1 data set to produce high SNR stacked spectra.

\section{Target Selection Accuracy}
Based on the extensive testing described in Section 4.2, it was
determined that the typical accuracy of the photometric redshifts
adopted in the VANDELS target selection was $\sigma_{\rm dz}\simeq
0.02$, with a catastrophic outlier rate of $\leq 2\%$. 
However, as is often the case, the samples of galaxies used to validate the photometric redshifts have $i-$band
magnitudes that are significantly brighter than those of the real VANDELS targets. 
Indeed, the median $i-$band magnitude of the galaxies used to validate
the photometric redshifts is two magnitudes brighter than the median
$i-$band magnitude of the DR1 galaxies. Consequently, it is
clearly of interest to use the DR1 galaxies to review the accuracy of the selection process based on real, on-sky, data.

In the top panel of Fig. 8 we show a plot of $z_{\rm phot}$ versus
$z_{\rm spec}$ for the galaxies released in DR1 with spectroscopic
redshift quality flags 3 and 4, which together comprise 55\% of the full DR1 sample.
For these galaxies $\sigma_{\rm dz}= 0.026$ with a catastrophic outlier rate of only 0.8\%. 
The middle panel in Fig. 8 is the equivalent plot for those DR1
galaxies with spectroscopic redshift quality flags 1, 2 and 9, which have $\sigma_{\rm dz}= 0.036$ and a catastrophic outlier rate of 3.6\%. 
Taken together, the full DR1 sample (i.e. flags $1-9$) has an accuracy
of $\sigma_{\rm dz}= 0.029$ with a catastrophic outlier rate of $2.1\%$.

It is worth noting that the fraction of catastrophic outliers is
actually significantly biased by the inclusion of a relatively small number of
AGN candidates and {\it Herschel-}detected galaxies.
If the statistics are restricted to the 97\% of objects drawn from the three principal
classifications of VANDELS targets (see Section 4.5), the accuracy is
$\sigma_{\rm dz}= 0.028$ and the catastrophic outlier rate is a
remarkably low 1.2\% (flags $1-9$). Given the relative faintness of the VANDELS targets, these figures 
provide a clear validation of the accuracy and robustness of the target selection procedure described in Section 4.
Moreover, the low number of catastrophic outliers amongst those
objects allocated spectroscopic quality flags 1 and 2 suggests that the VANDELS quality flags are somewhat conservative.
In reality, for many of the flag 1 and 2 objects we can be very confident that the
spectroscopic redshift lies within a relatively narrow range, but the 
spectral features simply do not allow competing redshift solutions to be reliably differentiated.

In the bottom panel of Fig. 8 the redshift distribution of the galaxies released in DR1 is shown as the filled blue histogram,
based on their measured spectroscopic redshifts. The histogram indicated
by the thin grey line shows the redshift distribution of the VANDELS
parent sample, based on the input photometric
redshifts. A comparison of the two clearly indicates that the spectroscopic
redshift distribution of the real VANDELS spectra is in very close
agreement to the distribution predicted by the photometric-redshift
selection procedure.

The galaxies targeted by the VANDELS survey are fainter
than those typically targeted by previous large spectroscopic surveys of high-redshift
galaxies. Consequently, it is clearly of interest to explore how the accuracy of
the VANDELS photometric redshifts varies as a function of target magnitude.

All but three of the VANDELS galaxies released in DR1 have $i-$band
magnitudes in the range $22.25 \leq i \leq 27.50$\footnote{One passive
  galaxy has $i=22.1$ and two further galaxies with $i\geq 27.5$ were selected as $z\geq 5.5$ LBGs based on their $z_{850}-$band magnitudes.}.
Consequently, Fig. 9 shows a comparison between spectroscopic and
photometric redshifts in three $i-$band magnitude ranges: $22.25 < i \leq 25.00$, $25.00<i
\leq 25.75$ and $25.75 < i \leq 27.50$, and includes all objects with
spectroscopic redshift quality flags $1-9$. The middle panel of Fig. 9
is representative of the $i-$band magnitude of the typical VANDELS
source, whereas the top and bottom panels illustrate the photometric
redshift accuracy at the bright and faint ends of the target magnitude
distribution, respectively. The relevant statistics quantifying the
quality of the agreement between the spectroscopic and photometric
redshifts are displayed in the top-left corner of each panel of Fig. 9.

It is clear from Fig. 9 that in terms of bias and catastrophic outlier rate, the VANDELS photometric redshifts perform very well within the two brighter magnitude bins.
Over the full magnitude range there is a gradual decrease in the photometric redshift accuracy, with $\sigma_{\rm dz}$ dropping from 0.025 to 0.036.
However, given the factor of $\simeq 5$ drop in brightness between the top and bottom panels, the decrease in accuracy is not particularly dramatic. 
In contrast, it is clear from the bottom panel of Fig. 9 that the
photometric redshifts for the faintest VANDELS targets with $i>25.75$
($\simeq 15\%$ of the DR1 objects) do show a notable increase in both the fraction of catastrophic outliers and the bias.

Overall, the quality of the VANDELS photometric redshifts is in-line with expectations
based on the spectroscopic redshift validation data (see Section 4.2). For all DR1 objects with spectroscopic
quality flags $1-9$, an accuracy of $\sigma_{\rm dz}=0.029$ and a
catastrophic outlier rate of 2.1\% compares favourably with the results from the
spectroscopic validation sets ($\sigma_{\rm dz}=0.025$ and $1.9\%$
catastrophic outliers), despite the $i-$band magnitudes of the VANDELS
galaxies being two magnitudes fainter than the validation objects, on average.
Interestingly, compared to the DR1 data, the overall systematic bias
of the photometric redshifts is only $0.003\pm0.002$. This is actually
better than the expectation from the spectroscopic validation data ($0.008\pm0.001$), albeit only at the $\simeq 2.5\sigma$ level.

\section{Summary and timeline}
In this paper we have provided an overview of the VANDELS 
spectroscopic survey, focusing on the scientific motivation, survey design and target selection.
The original motivation for the VANDELS survey was to move beyond
simple redshift determination and to provide the high SNR spectra necessary to study the physical properties of the high-redshift galaxy population.
The spectra released in DR1 demonstrate that the original goals of the
survey are within reach, and that the VIMOS spectrograph can be used
to integrate for 20--80 hours without the final SNR being dominated by systematic effects.
Combined with the unparalleled ancillary data available within the
CDFS and UDS survey fields, it is clear that the VANDELS survey has
the potential to become a key legacy data set for
studying the evolution of high-redshift galaxies for many years to come.

The observations for the VANDELS survey were fully completed in
February 2018. The second ESO public data release is currently
scheduled for June 2018 and will feature all of the spectra completed,
or 50\% completed, by the end of the second VANDELS observing season
in February 2017.  The third ESO public data release is scheduled for
June 2019 and will consist of the entire VANDELS spectroscopic data set.

A final data release is currently scheduled for June 2020 and will
formally mark the end of the project. It is currently intended that
the final data release will feature a re-reduction of the
entire spectroscopic data set, incorporating improvements in the data reduction
process which have been implemented over the course of the survey. 
In addition, the VANDELS team is committed to release two final catalogues
to enhance the legacy value of the survey. The first catalogue will contain
physical properties for each target (i.e. stellar masses, star-formation rates, dust attenuation and rest-frame colours) based on SED fitting of the final data set.
The second catalogue will provide measurements of the fluxes and equivalent
widths of significant emission/absorption features identified in the
VANDELS spectra, along with their corresponding uncertainties.

\section*{Acknowledgements}
Based on data products from observations made with ESO Telescopes at
the La Silla Paranal Observatory under programme ID 194.A-2003(E-K). 
We thank the ESO staff for their continuous support for the VANDELS survey, particularly the Paranal staff, who helped us to conduct
the observations, and the ESO user support group in Garching. RJM, AM, EMQ and DJM acknowledge funding from the European Research Council, via the award of an ERC
Consolidator Grant (P.I. R. McLure). AC acknowledges the grants PRIN-MIUR 2015 and ASI n.I/023/12/0. 
RA acknowledges support from the ERC Advanced Grant 695671
``QUENCH''.  F.B.  acknowledges the support by Funda\c{c}\~ao para a
Ci\^encia e a Tecnologia (FCT) via the postdoctoral fellowship
SFRH/BPD/103958/2014 and through the research grant UID/FIS/04434/2013.
PC acknowledges support from CONICYT through the project FONDECYT regular 1150216.

%%%%%%%%%%%%%%%%%%%%%%%%%%%%%%%%%%%%%%%%%%%%%%%%%%



\bibliographystyle{mnras}
\bibliography{vandels_sdef_submit_v2} % if your bibtex file is called example.bib

\bigskip

\noindent
$^{1}$Institute for Astronomy, University of Edinburgh, Royal Observatory, Edinburgh, EH9 3HJ, UK\\
$^{2}$INAF, Osservatorio Astronomico di Roma, Monteporzio, Italy\\
$^{3}$Dipartimento di Fisica e Astronomia, Universit\`{a} di Bologna, Via Gobetti 93/2, I-40129, Bologna, Italy\\
$^{4}$INAF - Osservatorio Astrofisico di Arcetri, Largo E. Fermi 5, I-50157, Firenze, Italy\\
$^{5}$Laboratoire AIM-Paris-Saclay, CEA/DRF/Irfu, CNRS France\\
$^{6}$MPE, Giessenbachstrasse 1, D-85748 Garching, Germany\\
$^{7}$Kavli Institute for Cosmology, University of Cambridge, Madingley Road, Cambridge CB3 0HA, UK\\
$^{8}$Cavendish Laboratory, University of Cambridge, 19 J. J. Thomson Avenue, Cambridge CB3 0HE, UK\\
$^{9}$INAF-Osservatorio di Astrofisica e Scienza dello Spazio di Bologna, via Gobetti 93/3, I-40129, Bologna, Italy\\
$^{10}$European Southern Observatory, Karl-Schwarzschild-Str. 2, D-85748 Garching b. Munchen, Germany\\
$^{11}$Observatoire de Gen\`{e}ve, Universit\'{e} de Gen\`{e}ve, 51 Ch. des Maillettes, 1290, Versoix, Switzerland\\
$^{12}$Department of Astronomy, The University of Texas at Austin, Austin, TX 78712, USA\\
$^{13}$INAF - Astronomical Observatory of Trieste, via G.B. Tiepolo 11, I-34143 Trieste, Italy\\
$^{14}$INAF-Istituto di Astrofisica Spaziale e Fisica Cosmica Milano, via Bassini 15, 20133, Milano, Italy\\
$^{15}$N\'ucleo de Astronom\'ia, Facultad de Ingenier\'ia, Universidad Diego Portales, Av. Ej\'ercito 441, Santiago, Chile\\
$^{16}$Department of Physics and Astronomy, University College London, Gower Street, London WC1E 6BT, UK\\
$^{17}$INAF-Osservatorio Astronomico di Brera, via Brera 28, 20122 Milano, Italy\\
$^{18}$Astrophysics, The Denys Wilkinson Building, University of Oxford, Keble Road, Oxford OX1 3RH, UK\\
$^{19}$Kapteyn Astronomical Institute, University of Groningen, Postbus 800, 9700 AV, Groningen, The Netherlands\\
$^{20}$European Southern Observatory, Avenida Alonso de C\'{o}rdova 3107, Vitacura, 19001 Casilla, Santiago de Chile, Chile\\
$^{21}$School of Physics and Astronomy, University of Nottingham, University Park, Nottingham NG7 2RD, UK\\ 
$^{22}$University Observatory Munich, Scheinerstrasse 1, D-81679 Munich, Germany\\
$^{23}$Department of Astronomy, University of Michigan, 311 West Hall, 1085 South University Ave., Ann Arbor, MI 48109-1107, USA\\
$^{24}$Instituto de Astrof\'{i}sica e Ci\^{e}ncias do Espa\c{c}o, Universidade de Lisboa, OAL, Tapada da Ajuda, P-1349-018 Lisbon, Portugal\\
$^{25}$Departamento  de  F\'{i}sica,  Faculdade  de  Ci\^{e}ncias, Universidade  de Lisboa, Edif\'{i}cio C8, Campo Grande, PT1749-016 Lisbon, Portugal\\
$^{26}$Instituto de Fisica y Astronomia, Facultad de Ciencias, Universidad de Valparaiso, 1111 Gran Bretana, Valparaiso, Chile\\
$^{27}$Institute  d'Astrophysique  de  Paris,  CNRS,  Universit\'{e} Pierre et Marie Curie, 98 bis Boulevard Arago, 75014, Paris, France\\
$^{28}$National Optical Astronomy Observatory, 950 North Cherry Ave, Tucson, AZ, 85719, USA\\
$^{29}$Harvard-Smithsonian Center for Astrophysics, 60 Garden St, Cambridge MA 20138, USA\\
$^{30}$Space Telescope Science Institute, 3700 San Martin Drive, Baltimore, MD, 21218, USA\\
$^{31}$Dark Cosmology Centre, Niels Bohr Institute, University of Copenhagen, Juliane Maries Vej 30, DK-2100 Copenhagen, Denmark\\
$^{32}$Imperial college, Kensington, London SW7 2AZ, UK\\
$^{33}$Astronomy Department, University of Massachusetts, Amherst, MA01003, USA\\
$^{34}$Pontificia Universidad Cat\'{o}lica de Chile, Instituto de Astrof\'{i}sica Avda. Vicu\~{n}a Mackenna 4860, Santiago, Chile\\ 
$^{35}$Aix Marseille Universit\'{e}, CNRS, LAM (Laboratoire d'Astrophysique de Marseille) UMR 7326, 13388, Marseille, France\\
$^{36}$Instituto de Astrof\'isica de Canarias, Calle V\'ia L\'actea s/n, E-38205 La Laguna, Tenerife, Spain\\
$^{37}$Departamento de Astrof\'isica, Universidad de La Laguna, E-38200 La Laguna, Tenerife, Spain\\
$^{38}$Faculty of Physics, Ludwig-Maximilians Universit\"{a}t, Scheinerstr. 1, 81679, Munich, Germany\\
$^{39}$Department of Physics and Astronomy, Texas A\&M University, College Station, TX 77843-4242, USA\\
$^{40}$Excellence Cluster, Boltzmann Strasse 2, D-85748 Garching, Germany\\
$^{41}$Department of Physics, Durham University, South Road, DH1 3LE Durham, UK\\
$^{42}$Leiden Observatory, Leiden University, 2300 RA, Leiden, The Netherlands\\ 
$^{43}$Steward Observatory, The University of Arizona, 933 N Cherry Ave, Tucson, AZ, 85721, USA\\
$^{44}$Department of Physics and Astronomy, PAB, 430 Portola Plaza, Box 951547, Los Angeles, CA 90095-1547, USA\\
$^{45}$School of Physics and Astronomy, University of St. Andrews, SUPA, North Haugh, KY16 9SS St. Andrews, UK
% Don't change these lines
\bsp	% typesetting comment
\label{lastpage}
\end{document}